\documentclass[prl,twocolumn,superscriptaddress,showpacs,amsmath,amssymb]{revtex4-2} 
\usepackage{graphicx}
\usepackage{amsmath}
\usepackage{dcolumn}
\usepackage{bm}
\usepackage[dvipsnames]{xcolor}
\usepackage{empheq}
\usepackage[makeroom]{cancel}
\usepackage[colorlinks,bookmarks=false,citecolor=olive,linkcolor=olive,urlcolor=olive]{hyperref}
\usepackage[normalem]{ulem}

\begin{document}

\begin{abstract}  Confining light around solids via cavities enhances the coupling between the electromagnetic fluctuations and the matter. We predict that in superconductors this cavity-enhanced coupling enables the control of the order-parameter stiffness, which governs key length scales such as the coherence length of Cooper pairs and the magnetic penetration depth. We explain this as a renormalization of the Cooper-pair kinetic mass caused by photon-mediated repulsive interactions between the electrons building the pair. This effect is generic for Bardeen-Cooper-Schriffer superconductors and is most pronounced in low-$T_c$ materials. The strength of this effect can be tuned via the length of the cavity and we estimate it to be sizable for cavities in the infrared range. 
\end{abstract}

\title{Cavity-control of the Ginzburg-Landau stiffness in superconductors}
\author{Vadim Plastovets}
\author{Francesco Piazza}
\affiliation{Theoretical Physics III, Center for Electronic Correlations and Magnetism, Institute of Physics, University of Augsburg, 86135 Augsburg, Germany}
\date{\today}
\maketitle

\textit{Introduction}.-- 
Geometrical confinement of the vacuum photons in optical cavities is known to produce strong light-matter interactions \cite{frisk2019ultrastrong}. As a result, low-energy photons mediate forward electron scattering, which provides a control over stationary states (like the ground or thermal equilibrium state) opposed to the transient regime available within laser-based approaches. The fundamentally new ingredient in cavity-quantum-materials is that the electromagnetic (EM) field is an active degree of freedom with an intrinsic quantum nature, as opposed to an external classical field unaffected by what happens in the material. This can have a profound impact and the emerging field of cavity quantum materials has featured promising developments in recent years \cite{Garcia-Vidal:2021, Mivehvar:2021, Schlawin:2022, Bloch:2022}. 
In particular, experiments have demonstrated the possibility of affecting the macroscopic phase of a material using cavities without any drive. Examples involve quantum Hall phases \cite{appugliese2022breakdown}, ferromagnetism \cite{Thomas:2021}, charge-density-waves \cite{Jarc:2023}, and also promising steps in the control of superconductors \cite{Thomas:2019,Keren2026,xu2026}.

The advancements in the theory of cavity-superconductor physics can be divided into two main areas. The first deals with the microscopic origin of the photon-mediated pairing between electrons and the resulting superconductivity \cite{sentef2018cavity,PhysRevLett.122.133602,chakraborty2021long,PhysRevB.109.104513,eckhardt2024theory,lu2024cavity,kozin2025cavity}. The second studies the interplay between the low-energy collective excitations of the superconductor (Josephson plasma waves, Higgs and Bardasis-Schrieffer, Majorana modes) and the cavity photonic mode \cite{PhysRevB.93.075152,PhysRevB.99.020504,PhysRevResearch.2.013143,PhysRevB.104.L140503,dmytruk2024hybrid,bacciconi2024topological}. Related work has been done also dealing with ultracold atomic gases in cavities \cite{ZwettlerBrantut2025,Schlawin2019atoms,ortuno2025pauli,frank2025fate}.

In the following we instead explore how the cavity vacuum field can be used to control the emergent length scales characterizing the superconducting phase. We systematically derive the Ginzburg-Landau (GL) equation including the effect of the cavity-controlled coupling between the vacuum EM field and the electrons. We show that, by tuning the cavity length, the superconductor coherence length and London penetration depth $\lambda_L$ can be widely controlled under realistic conditions. For conventional low-$T_c$ materials, such as Al or Nb, we namely predict an order of magnitude increase in the London penetration depth. Differently from conventional methods, such as classical laser-induced local thermal suppression or non-thermal interactions via intense THz radiation \cite{veshchunov2016optical,sekiguchi2024anomalous,de2025generation}, which often involve transient dynamics or pair-breaking, this approach offers a new, non invasive pathway for optical controlling superconductivity at thermal equilibrium. 
%

\begin{figure}[] 
\begin{minipage}[h]{1.0\linewidth}
\includegraphics[width=1.0\textwidth]{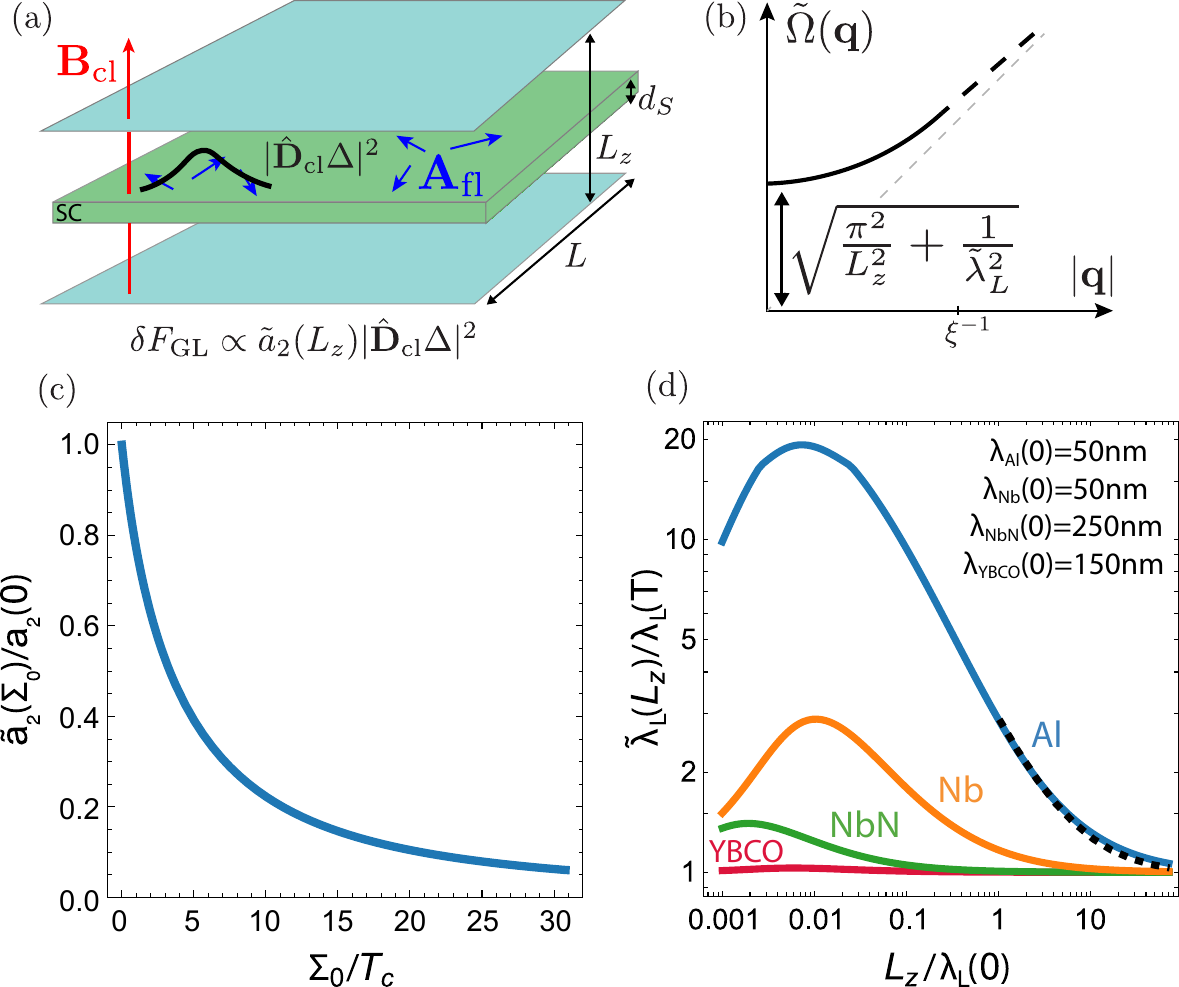} 
\end{minipage}
\caption{\small{
(a) Sketch of a superconductor (green) inside a Fabry-Pérot cavity (grey). "Deformation" of the order parameter by the operator $\hat {\bf D}_\text{c{}l}=-i\nabla-2e{\bf A}_\text{c{}l}$ contributes to the GL free energy $\delta F_\text{GL}$ via the GL stiffness parameter $\tilde a_2$, which depends on cavity size. (b) Gapping of the photon field spectrum due to cavity confinement and the Meissner effect (for $\tilde\lambda_\text{L} \lesssim d_\text{S}$) inside the superconductor. (c)  $\tilde{a}_2$ versus electron-photon interaction energy $\Sigma_0$ with bare value $a_2(0) = \nu_{2D}\xi^2(0)7 \zeta(3)/8$. (d) Renormalized magnetic penetration depth $\tilde\lambda_\text{L}$ as a function of cavity length $L_z$ for four typical materials with zero-temperature values $\lambda_\text{L}(T=0)$. Black dashed line shows self-consistent Meissner renormalization effect for Al at $T/T_c=0.8$. }}
\label{fig_1}
\end{figure}

\textit{Classical vs quantum electromagnetic field}.-- 
We consider a simple setup of a quasi two-dimensional BCS superconducting film with thickness $d_\text{S}$, positioned in the center of a Fabry-Pérot cavity of the size $L^2\times L_z$, as illustrated in Fig. \ref{fig_1}(a). This system is modeled by electrons interacting via conventional short-range BCS attraction and coupled to the EM field, assumed to be in thermodynamic equilibrium with the superconductor. The partition function $\mathcal{Z}$ of the system can be defined as a path integral with the corresponding Euclidean action in imaginary time \cite{VanOtterlo_1999} 
\begin{gather}\label{S_tot}
    S[\psi, {\bf A}] =  \int d{\bf r}d{\tau} ~ \bar{\psi}_\sigma \left[\partial_\tau + \hat{\xi}(\nabla-ie{\bf A})  \right] {\psi}_\sigma 
    \\ \notag
    - \lambda\bar{\psi}_\uparrow\bar{\psi}_\downarrow \psi_\downarrow\psi_\uparrow + \frac{{\bf E}^2+{\bf B}^2}{8\pi}.
\end{gather}
Here $\psi_\sigma({\bf r}, \tau)$ is a Grassman variable describing electrons; $\hat{\xi}(\nabla)=\frac{1}{2m}(-i\nabla)^2-\mu$ is the kinetic energy operator; $\lambda$ is the BCS coupling constant; $e$ is electron charge; ${\bf E}=\partial_\tau {\bf A}$, ${\bf B}=\nabla \times {\bf A}$ are the components of the EM field determined by the vector potential ${\bf A}({\bf r},\tau)$, while the scalar potential is gauged to zero. Hereafter we use natural units $\hbar=c=k_B=1$.

We split the EM vector potential as ${\bf A}={\bf A}_\text{cl}+{\bf A}_\text{fl}$, where ${\bf A}_\text{cl}$ is the \textit{deterministic} part satisfying classical Maxwell's equations ($\delta S/\delta {\bf A}|_{{\bf A}={\bf A}_{\rm cl}}=0$), and ${\bf A}_\text{fl}$ represents fluctuations of the EM field. ${\bf A}_\text{fl}$ includes both  \textit{thermal} (classical, dominant at finite temperature) and \textit{quantum} (zero-point motion) fluctuations. While thermal fluctuations can strongly affect the ground state in free space \cite{PhysRevLett.32.292}, they are suppressed in the present-cavity system due to the creation of a photonic gap (see End Matter). We thus focus on the quantum fluctuations of the EM field in the following. From the GL theory perspective, these quantum fluctuations are fast dynamic modes that cannot couple directly to the superconducting order parameter and must be integrated out at the electronic level. Therefore, we proceed as follows: (i) integrate out the EM fluctuations to obtain a photon-mediated interaction between electronic currents (so-called Amperean), which we treat perturbatively; (ii) apply mean-field decoupling of the BCS interaction, then integrate out electrons to derive a GL effective action for the order parameter $\Delta({\bf r})$. As we shall see, we obtain only negligible corrections to electron kinetics and the superconducting critical temperature, but appreciable modification of the GL stiffness parameter, allowing to control superconducting properties via cavity geometry.

\textit{Interaction between electrons mediated by the electromagnetic fluctuations}.-- 
The geometrical confinement from the cavity imposes boundary conditions at the cavity mirrors for the EM field, which lead to a gap in the dispersion relation for the photons.
In the Coulomb gauge $\nabla \cdot {\bf A}=0$, the transverse part of the EM field in the middle of the cavity at $z=L_z/2$ is described by the following vector potential: ${{\bf A}_\text{f{}l}({\bf r}_{||},\tau) =\sqrt{\frac{2}{VT}} \sum_{q,s} e^{i({\bf r}_{||}{\bf q}+\Omega_m\tau)}{\bf e}_{{\bf q},s}\mathcal{A}_s({\bf q},i\Omega_m)}$, where $\Omega_m$ is bosonic Matsubara frequency, ${\bf q}$ is in-plane photon momentum and $\sum_q=T\sum_{\Omega_m,{\bf q}}$. 
We adopt the single longitudinal-mode approximation, since for submicron cavity lengths the mode spacing $\Omega_0=\pi/L_z$ greatly exceeds all relevant electronic energy scales in the normal state \cite{PhysRevB.104.235120}, and remains much larger than the superconducting gap in the condensed phase, where $T_c \ll \Omega_0$. The vector field is written in the basis of TE/TM ($s=1,2$) polarizations with the unit vectors ${\bf e}_{{\bf q},1} = {\bf q}/|{\bf q}|$ and ${\bf e}_{{\bf q},2} = {\bf e}_{{\bf q},1}\times {\bf z}_0$, which form orthonormal basis ${\bf e}_{{\bf q},s}{\bf e}_{{\bf q}',s'}=\delta_{ss'}$ \cite{PhysRevA.50.1830}. 

The propagation of transverse photons inside a superconductor is in general modified by screening effects. However, in the following regime: ${d_\text{S} \ll \lambda_\text{L} \ll \Lambda_\text{P} \sim L}$, where $\lambda_\text{L}$ is the London penetration depth and ${\Lambda_\text{P} = 2\lambda_\text{L}^2/d_\text{S}}$ is the Pearl penetration depth for thin films \cite{10.1063/1.1754056}, one can safely neglect the screening and assume the homogeneity of the transverse EM field inside the superconductor. We discuss the interplay between cavity and screening effects later. After integrating out the photonic degrees of freedom we obtain an effective action 
\begin{gather}\label{S_el_ph}
    S_\text{el-pht}[\psi] =  \sum
    _{\substack{k,k',{\bf q} \\ \sigma,\sigma'}}
    \frac{V_{{\bf k},{\bf k}'}({\bf q})}{2}
    \bar\psi_{k-{\bf q} \sigma}\psi_{{k} \sigma}\bar\psi_{{k}'+{\bf q} \sigma'}\psi_{{k}' \sigma'},
\end{gather}
where $\psi_{{k} \sigma}$ the Fourier transform of the fermionic field $\psi({\bf r}_{||},\tau)$, and we used the notation $k=\omega_n,\bf{k}$ with the fermionic Matsubara frequency $\omega_n$. 
The photon-mediated static interaction potential reads 
\begin{gather}\label{V_dyn}
    V_{{\bf k},{\bf k}'}({\bf q}) 
    = -V_0  \frac{ ({\bf k}-\frac{\bf q}{2}) ({\bf k}'+\frac{\bf q}{2}) }{\Omega^2({\bf q})}
\end{gather}
with the amplitude $V_0=\frac{8\pi e^2}{V m^2}$ and the photon spectrum $\Omega^2({\bf q})= \Omega_0^2 + |{\bf q}|^2$.

%

\begin{figure}[] 
\begin{minipage}[h]{1.0\linewidth}
\includegraphics[width=0.95\textwidth]{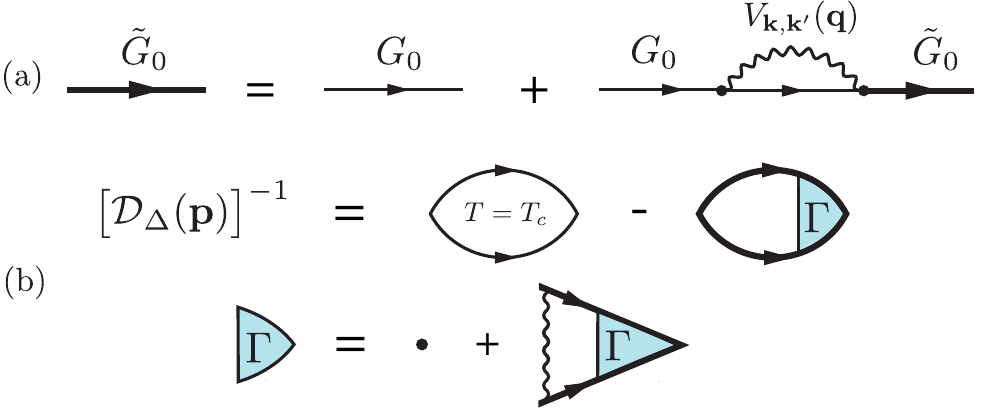} 
\end{minipage}
\caption{\small{ (a) Dyson equation for the Green function $\tilde{G}_0$ dressed by the effective electron-photon interaction vertex $V_{{\bf k, k'}}({\bf q})$ via the self-energy $\Sigma({\bf k})$. (b) Inverse propagator of the gap field ${\big[ \mathcal{D}_\Delta({\bf p}) \big]^{-1} =  \Pi(0)|_{T_c} - \tilde\Pi({\bf p})}$ with the bare BCS interaction $\lambda^{-1}=\Pi(0)|_{T_c}$ and the vertex corrections $\Gamma({\bf k},{\bf p})$. 
}}
\label{fig_2}
\end{figure}

\textit{Modification of the kinetics of electrons}.-- 
The photon-mediated interaction \eqref{S_el_ph}  renormalizes the free electron propagator $G^{-1}_0({\bf k},i\omega_n)=-i\omega_n+\xi_{\bf k}$ to ${\tilde{G}_0^{-1}({\bf k},i\omega_n) =  G_0^{-1}({\bf k},i\omega_n) - \Sigma({\bf k})}$, which is represented diagrammatically in Fig. {\ref{fig_2}}(a). Hereafter the tilde sign denotes cavity-renormalized quantities. To first order in $V_0$, the Fock-like contribution to the self-energy reads as \cite{altland2010condensed} ${\Sigma({\bf k}) = \sum_{{k}'}V_{{\bf k, k'}}({\bf k}-{\bf k}') G_0({\bf k}',i\omega_n')}$. Note that, in principle, the self-energy should be computed including the classical EM field $\bf{A}_\text{cl}$ in the electron propagator $G_0$. We will however neglect this interference between the EM fluctuations and the classical EM field, and include the latter only later in the GL action. The momentum summation in $\Sigma({\bf k})$ can be performed analytically using the forward-scattering nature of $V_{{\bf k},{\bf k}'}({\bf q})$ i.e. approximating it with a delta function around ${\bf q} = 0$. We obtain the following isotropic contribution $\Sigma(k) \approx \Sigma_0 (|{\bf k}|^2/k_F^2)n_F(\xi_k)$, where $ \Sigma_0 = 32 c_0 \alpha E_F^2/k^2_F L_z$ is the amplitude of the self-energy, $n_F$ is Fermi-Dirac distribution function, $\alpha$ is the fine-structure constant, $E_F$ is the Fermi-energy and ${c_0=2\pi \ln(k_F/\Omega_0)}$. 

The main momentum dependence of the self-energy comes from the thermal distribution function and it affects mostly the region around $k_F$. For example, the linearized quasiparticle spectrum close to the Fermi surface ${\tilde \xi_k=\pm \tilde v_F (k-k_F)}$ attains a renormalized Fermi velocity ${\tilde v_F = v_F - \partial_k \Sigma(k)|_{k_F}}$. Unless one considers deep subwavelength cavities, these effects are small due to the factor $\propto 1/(k_F L_z)$ in $\Sigma_0$, as discussed previously in the literature \cite{PhysRevLett.122.133602,PhysRevB.104.235120,PhysRevB.105.165121}. As we shall see below, the kinetics of Cooper pairs can be instead appreciably affected by the photon-mediated interaction between electrons, as it is governed by the emergent energy scale set by the superconducting critical temperature.

\textit{Modification of the kinetics of Cooper pairs}.-- 
After its effect on the electrons, we want now  to consider the effect of the photon-mediated interaction \eqref{S_el_ph} onto Cooper pairs. As anticipated, we do not consider the Amperean (momentum-dependent) corrections to the coupling constant $\lambda$ \cite{PhysRevLett.122.133602}, assuming the phonon mechanism to be dominant. 
We start by rewriting the $s$-wave BCS interaction in Fourier space as
\begin{gather}\label{S_bcs}
    S_\text{BCS}[\psi] = - \frac{\lambda}{T} \sum_{k, k', {\bf p}}   
    \bar\psi_{k+\frac{{\bf p}}{2}\uparrow}\bar\psi_{-k+\frac{{\bf p}}{2}\downarrow}\psi_{-k'+\frac{{\bf p}}{2}\downarrow}\psi_{k'+\frac{{\bf p}}{2}\uparrow} 
\end{gather}
We proceed with the usual mean-field decoupling of the BCS action \cite{altland2010condensed}, which is achieved via a Hubbard-Stratonovich transformation involving an additional path-integral over the pairing-gap field $\Delta({\bf p})= \lambda \sum_k \langle \psi_{-k+{\bf p}/2\downarrow} \psi_{k+{\bf p}/2\uparrow} \rangle$. After integrating out the electronic degree of freedom, the effective action for $\Delta({\bf p})$ close to the critical temperature $T_c$ is Gaussian to leading order: ${S_\text{eff}[\Delta] \overset{T\rightarrow T_c}{=}  \sum_{\bf p} \bar\Delta({\bf p}) \big[ \mathcal{D}_\Delta({\bf p}) \big]^{-1} \Delta({\bf p})}$, where the inverse gap propagator reads ${\big[ \mathcal{D}_\Delta({\bf p}) \big]^{-1} =  \lambda^{-1} - \tilde\Pi({\bf p})}$, and $\lambda^{-1} =  \Pi(0)|_{T_c}$ defines the bare superconducting transition temperature $T_c$.

The gap propagator $\mathcal{D}_\Delta({\bf p})$ contains the polarization function $\tilde\Pi({\bf p})$,  describing the response of electrons to a pairing field through creation and annihilation of Cooper pairs (see Fig. \ref{fig_2}(b)). Such processes are at the same time subject to the additional photon-mediated interaction \eqref{S_el_ph}. In a perturbative treatment of the latter, the leading order contribution to the polarization function $\tilde\Pi({\bf p})$ is the so-called vertex correction $\Gamma({\bf k},{\bf p})$ (graphically this is shown in Fig. \ref{fig_2}(b)):
\begin{gather}\label{Pi_q_tau}
    \tilde\Pi ({\bf p}) = \sum_{k} \tilde G_0({\bf k_+},i\omega_n)\tilde G_0({\bf -k_-},-i\omega_n)\Gamma({\bf k},{\bf p}).
\end{gather}
The irreducible part of $\Gamma({\bf k},{\bf p})$ 
includes the same interaction potential as the self-energy, namely $\delta\Sigma({\bf k})/\delta G_0({\bf k}') = V_{{\bf k},{\bf k}'}({\bf k}-{\bf k}')$. We can sum the multiple scattering events within the ladder approximation, so that the $\Gamma$ satisfies the self-consistent Bethe-Salpeter equation:
\begin{gather}\label{BSeq}
    \Gamma({\bf k},{\bf p}) = 1 + T\sum_{{\bf k}',n} V_{{\bf k}_+,-{\bf k}_-}({\bf k}-{\bf k}')\Gamma({\bf k}',{\bf p}) 
    \\ \notag \times \tilde G_0({\bf k}'_+,i\omega_n)\tilde G_0({\bf -k}'_-,-i\omega_n),
\end{gather}
where ${\bf k}_{\pm}={\bf k}\pm{\bf p}/2$. Using the same forward-scattering approximation as for $\Sigma({\bf k})$, we can obtain an explicit expression of the full vertex ${\Gamma({\bf k},{\bf p}) = \left[1+\mathcal{F}({\bf k},{\bf p})\right]^{-1}}$, where ${\mathcal{F} = \Sigma_0 ({\bf k}_+{\bf k}_-/k_F^2) T\sum_n \tilde G_0({\bf k_+},i\omega_n)\tilde G_0({\bf -k_-},-i\omega_n)}$.

We are interested in the gradient term of the GL expansion of the effective action, which is obtained by the expansion of the polarizaton operator for small momenta ${|{\bf p}|\sim \xi^{-1}(T)\ll\xi^{-1}(0)\ll k_F}$, where we introduced the superconducting coherence length in 2D ${\xi(T)=\sqrt{7\zeta(3)/8}\xi(0)(1-T/T_c})^{-1/2}$ with $\xi(0)=v_F/2\pi T_c$. Finally, the gap-field propagator can be written as 
\begin{gather}\label{F_0}
    \big[ \mathcal{D}_\Delta({\bf p}) \big]^{-1} = a_0 + \tilde{a}_2(\Sigma_0){\bf p}^2 + \mathcal{O}({\bf p}^4),
\end{gather}
where $a_0=\Pi(0)|_{T_c} -  \Pi(0)\approx \nu_{2D} \ln(T/T_c)$, with the two-dimensional density of states $\nu_{2D} = m/2\pi$. We assume $\tilde{\Pi}(0) = \Pi(0)$, since as stated previously we neglect the correction induced by the photon-mediated interaction to $\lambda$ and thus to $T_c$. 

The vertex correction to the polarization function, induced by the electromagnetic fluctuations in the cavity, results in a renormalization of the GL stiffness parameter $\tilde{a}_2 = -\frac12 \partial^2_{\bf p} \tilde\Pi({\bf p})|_{{\bf p}=0}$, which depends on the ratio between the effective electron-photon interaction strength $\Sigma_0$ and superconducting critical temperature $T_c$. The numerical computation of $\tilde{a}_2(\Sigma_0)$ is shown in Fig. \ref{fig_1}(b). The behaviour or $\tilde{a}_2(\Sigma_0)$ can be understood quite intuitively. The Amperean interaction \eqref{V_dyn} is repulsive between the electrons in a Cooper pair, but it becomes less repulsive for larger center-of-mass momenta of the pair. This implies that the effect of these photon-mediated interactions is to increase the kinetic mass of the Cooper pair: $m_\text{kin}\propto \tilde a_2^{-1}$. A heavier pair, in turn, shortens the superconducting coherence length as follows
\begin{gather}\label{tilde_xi}
    \tilde\xi^2(\Sigma_0, T) =  \xi^2(T) ~ [\tilde{a}_2(\Sigma_0)/\tilde{a}_2(0)].
\end{gather}

So far, we have omitted the classical component of the EM field, which represents both the external probe and the field induced by the internal superconducting currents. The classical field must be incorporated into the electron propagator $\tilde{G}_0$ entering Eq. \eqref{Pi_q_tau}. Within the local GL approximation, where the EM field carries zero momentum, this approach is equivalent to the gauge-invariant formulation using the covariant derivative of the gap field, $\hat{\bf D}_\text{cl}=-i\nabla - 2e{\bf A}_\text{cl}$, yielding the effective action
\begin{gather}\label{F_GL}
    S_\text{GL}[\Delta,{\bf A}_\text{cl}]
    = \int d{\bf r} \Big[ a_0 |\Delta|^2 + \tilde{a}_2\big|\hat{\bf D}_\text{cl} \Delta\big|^2 + \frac{{\bf B}^2_\text{cl}}{8\pi} \Big].
\end{gather}
As anticipated, in the regime of validity of our model, the quantum EM fluctuations affect the above GL free energy only through parameter renormalization 
\footnote{This contrasts with the effect of purely thermal fluctuations studied in Ref. \cite{PhysRevLett.32.292}. There, the transverse EM field acquires a gap induced by the Meissner effect and couples directly to the mean-field $\Delta$ in the  GL action. Since the Meissner gap vanishes near $T_c$, these fluctuations become very strong and, once integrated out, reshape the free energy profile and alter the nature of the phase transition. This effect is intrinsic and observable in type-I superconductors, but in our case it is rendered negligible by the cavity-induced photonic gap, that remains finite also at $T_c$}
.

The stiffness parameter $\tilde{a}_2(\Sigma_0)$ describes the renormalization of both superconducting coherence length and the magnetic penetration depth
\begin{gather}\label{tilde_lambda}
    \tilde \lambda_\text{L}^2(\Sigma_0, T) = \lambda_\text{L}^2(T) ~ [\tilde{a}_2(0)/\tilde{a}_2(\Sigma_0)],
\end{gather}
which is shown in Fig. \ref{fig_1}(d).
Close to $T_c$ and for spatially uniform $|\Delta|$, we can express ${\tilde \lambda_\text{L}^2 = m/4\pi e^2 \tilde n_s}$ via the superfluid density $\tilde n_s(\Sigma_0, T)=8m\tilde a_2\Delta_0^2(T)$, and we observe that the cavity leads to the suppression of the latter. Moreover, controlling $\tilde a_2(\Sigma_0)$ also means that we can tune the GL parameter, defined as $ \tilde \kappa_\text{GL}(\Sigma_0) = \tilde \lambda_\text{L}(\Sigma_0, T)/\tilde\xi(\Sigma_0, T)$. This in turn implies that we can use the cavity to drive the system through the crossover between type-I ($\tilde \kappa_\text{GL}\lesssim 1/\sqrt2$) and type-II ($\tilde \kappa_\text{GL}\gg 1/\sqrt2$) superconductivity. As shown in Fig. \ref{fig_1}(d) and discussed in more detail later, this can be achieved by tuning the cavity length.

\textit{Screening problem}.-- 
The cavity-size dependence of the London penetration depth $\tilde\lambda_\text{L}$ leads to a self-consistent screening problem. To isolate the transverse electrostatic response, we assume $|\Delta|$ to be constant and set longitudinal fields and the order parameter phase to zero. The resulting action reads as 
${S_\text{ph}[{\bf A}] = \frac{1}{2}T \sum_{{\bf q}} {\bf A}(-{\bf q})\mathcal{D}^{-1}({\bf q}){\bf A}({\bf q})}$, where 
the propagator in the long-range (Pearl) regime: $\tilde\lambda_\text{L} \gg d_\text{S}$ is ${\mathcal{D}^{-1} = 2(\tilde\lambda_\text{L}^2/d_\text{S})^{-1}+2|{\bf q}|}$ \cite{10.1063/1.1754056}; and for the local
\footnote{Strictly speaking, one must account for the more general and nonlocal Pippard regime $D^{-1}_\text{cl}({\bf q}) = Q({\bf q})+{\bf q}^2$, which exists for $\tilde \kappa_\text{GL}(\Sigma_0) \lesssim 1/\sqrt{2}$. However, in the dirty case with small scattering length $\ell\ll \xi$ the Pippard kernel $Q({\bf q}) \rightarrow \tilde{\lambda}_L^{-2}$ and we restore the local electrodynamics.}
(London) regime: $\tilde\lambda_\text{L} \lesssim d_\text{S}$ it becomes ${\mathcal{D}^{-1} =\tilde{\lambda}_L^{-2}+{\bf q}^2}$. 

When the superconductor is initially in the thin-film (Pearl) limit $d_\text{S}\ll\lambda_\text{L}\ll\Lambda_\text{P}\sim L$, the Meissner screening of the quantum fluctuations is negligible, and ss the cavity confinement leads to an even larger $\tilde\lambda_\text{L}$, the Meissner effect can be further suppressed. In contrast, in the thick-film (London) limit where $d_\text{S}\sim\lambda_\text{L}\ll L$, the photon dispersion acquires the additional mass term: $\Omega^2({\bf q}) \rightarrow \tilde\Omega^2({\bf q}) = ( \Omega_0^2 + \tilde\lambda_\text{L}^{-2} )+{\bf q}^2$ [see Fig. \ref{fig_1}(b)]. This shift enters the self-energy $\tilde\Sigma_0$ via the coefficient $c_0$, and one can estimate the self-renormalization of $\tilde\lambda_\text{L}$ via the simple relation $\tilde \lambda_\text{L}^2(\tilde\Sigma_0)/\lambda_\text{L}^2(0) =  \tilde{a}_2(0)/\tilde{a}_2(\tilde\Sigma_0)$. The resulting self-consistent correction is however logarithmically small even when $L_z\gg\tilde\lambda_\text{L}$, and the photon gap induced by the cavity mirrors remains dominant. The self-consistent shift of $\tilde\lambda_\text{L}$ is plotted in Fig. \ref{fig_1}(d).

Finally, we emphasize that the GL framework is inherently local and breaks down for type-I superconductors in the presence of finite-momentum electromagnetic fields. However, in thin films (typically dirty due to edge scattering) the coherence length is reduced to $\xi_d = \sqrt{\xi \ell}$ with the mean-free path $\ell \ll \xi$. This suppression allows the system to appear type-II even for small GL parameters $\tilde\kappa_\text{GL} \lesssim 1/\sqrt{2}$, effectively masking the type-I to type-II crossover.

\textit{Experimental observability}.-- 
To quantitatively probe the cavity-induced renormalization of the GL stiffness, one can either extract the upper critical field $H_{c2}=\Phi_0/2\pi\tilde\xi^2(L_z)$ from transport measurements \cite{robbins2025upper}, or characterize the Meissner screening via THz transmission/reflection spectroscopy \cite{Jarc:2023,10.1063/5.0080045}. As we shall see next, the relevant range for the cavity's resonant frequency is $\Omega_0 \sim 10^2$-$10^3$ THz, suggesting that standard THz probe experiments through the cavity are feasible within the proposed setup. Another approach can be a direct measurements of the London depth $\tilde{\lambda}_L(L_z)$ with the help of a magnetic atomic force microscope. This method has been used in Ref. \cite{Keren2026}, where cavity-induced suppression of the superfluid density has been observed.

The dimensionless ratio quantifying the amount of the cavity-controlled renormalization of the superconducting stiffness is
\begin{gather}\label{dimensionless_coupling}
    \Sigma_0/T_c = \alpha 16 \pi \ln\left(1+k_F^2L_z^2/\pi^2 \right) [E_F/T_c] [\lambda_C/L_z],
\end{gather}
where $\alpha = 4\pi e^2=1/137$ is QED fine structure constant and $\lambda_C=\hbar/m^*c$ is the Compton wavelength associated with the effective electron mass $m^*$ in the material. The non-monotonous dependence of $\Sigma_0$ on the cavity length leads to the non-monotonous behavior of $\tilde\lambda_\text{L}(L_z)$ observed in Fig \ref{fig_1}(d). We see that Eq. \eqref{dimensionless_coupling} contains the ratio $E_F/T_c$. As anticipated, it appears because the GL stiffness features $T_c$ as an additional intrinsic scale, which obviously does not appear when considering the impact of the photon-mediated interaction on the critical temperature itself. While the small ratio $\lambda_C/L_z$ typically makes the impact of photon-mediated Amperean interactions on $T_c$ negligible \cite{PhysRevLett.122.133602,PhysRevB.109.104513}, we find that it can nonetheless appreciably impact the superconducting stiffness, provided $E_F/T_c$ is sufficiently large. This  makes low-$T_c$ superconductors particularly favorable for our purposes. For the latter we usually have $m^*\approx m_e$, and thus $\lambda_C \approx 2.4$ pm. Given the condition $L_zk_F\gg 1$, we can estimate $c_0 = 2\pi \ln\left(k_FL_z/\pi\right)\sim 20$ for the entire range of $L_z$ and get $\Sigma_0/T_c \approx 5 (E_F/T_c)(\lambda_C/L_z)$. More precisely, for Al with $T_c = 1.2$ K and ${E_F/T_c = 9\times 10^4}$ we get $\Sigma_0/T_c \approx 1.08 L_z^{-1}[\mu \text{m}]$, and for YBCO with $T_c \approx 90$ K and ${E_F/T_c = 1.3\times 10^2}$ we get $\Sigma_0/T_c \approx 0.0015 L_z^{-1}[\mu \text{m}]$. 

Remarkably, measurement with NbN placed inside a cavity have just been performed \cite{xu2026}, showing indeed a significant suppression of the superfluid density $n_s$. Since $n_s(T\to T_c)\propto \lambda_L^{-2}$, these measurements can be explained by the mechanism we investigate in this work (see orange line in Fig. \ref{fig_1}). Importantly, the fact that the experiment did not observe appreciable changes in the gap is a strong indication in favor of the present mechanism, which relies solely on the dependence of the Cooper-pair kinetic energy on the cavity size, rather than a mechanism based on the modification of the phonon-mediated interaction responsible for superconductivity \cite{sentef2018cavity,PhysRevLett.122.133602,chakraborty2021long,PhysRevB.109.104513,eckhardt2024theory,lu2024cavity,kozin2025cavity,Keren2026}. Moreover, the latter mechanism requires tuning the cavity frequency close to a phonon mode, while our mechanism does not need such a resonant condition and could be observed in a broad range of conventional superconducting compounds.

\textit{Discussion}.-- 
We predicted that the Amperean interaction between electrons mediated by vacuum fluctuations inside a cavity can significantly alter the kinetics of Cooper pairs, and in turn the GL stiffness of a superconductor at fixed temperature and gap. This allows for non-invasive control of the  magnetic London penetration depth and coherence length by tuning the longitudinal size of a Fabry-Pérot cavity, enabling the material to be driven deeper into the type-II regime with enhanced $H_{c2}$ field, while also confining spatial variations of the order parameter, which could potentially improve scalability in superconducting microelectronic circuits \cite{10.1063/1.4948618}. Low-$T_c$ materials should be  best suited, and we expect a sizable growth of the London depth [see Fig. \ref{fig_1}(d)] to be reachable in infrared cavities.

In the future, it would be interesting to explore the possibility to produce the opposite effect; namely, an enhancement of the screening through an increase in $\tilde{n}_s$, which would also have promising applications, for instance improving the performance of thin-film Superconducting Quantum Interference Devices (SQUIDs). One possible approach would be to replace the current (Amperean) interaction between electrons, which is always repulsive in the Cooper $s$-wave channel, with a laser-assisted cavity-mediated interaction as proposed for instance in \cite{gao2020photoinduced,chiocchetta2021cavity}. With the proper choice of the laser-cavity detuning this interaction is attractive between electrons in a Cooper pair, and should thus reduce the kinetic mass of the Cooper pair through the same mechanism discussed in this work, thereby increasing the superfluid density.

\textit{Acknowledgments}.-- 
We thank Michele Pini for valuable comments. This research is supported by the Staatsministerium für Wissenschaft und Kunst through the Hightech Agenda Bayern Plus and is part of the Munich Quantum Valley and received additional support from the Deutsche Forschungsgemeinschaft (DFG, German Research Foundation) under the Walter Benjamin Programme (Project No. 566401345).

\newpage

\begin{widetext}
    \center{\textbf{End Matter}}
\end{widetext}

\setcounter{equation}{0}
\renewcommand{\theequation}{A\arabic{equation}}

\textit{Appendix A: Photon propagator}.-- 
The schematic derivation of the transverse photon propagator in a superconductor is as follows. After integrating out electronic degrees of freedom, the photonic part of the action \eqref{S_tot} acquires the mean-field form ${S_\text{ph}[\Delta, {\bf A}] =  \int d{\bf r}d{\tau} ~ ({\bf E}^2+{\bf B}^2)/(8\pi) - {\bf j}_s {\bf A}}$. The linear response of the supercurrent to the transverse field ${\bf j}_s({\bf q}) = - Q_\text{M}({\bf q},T){\bf A}({\bf q})$ encodes the Meissner effect and leads to a quadratic photon action ${S_\text{ph} \sim \sum_q {\bf A}(-q)\mathcal{D}^{-1}(q){\bf A}(q)}$ with ${\mathcal{D}^{-1} = \Omega_m^2+\Omega^2({\bf q})+4 \pi Q_\text{M}({\bf q},T)}$ in momentum/Matsubara representation $q={\bf q}, \Omega_m$. Here $\Omega({\bf q})$ is the dispersion induced by cavity geometry. The classical (deterministic) field corresponds to $\Omega_m=0$ mode, which becomes dominant in the high-temperature limit, and it is defined as a saddle-point of the photonic action $\delta S_\text{ph}/\delta A(-q,0)\big|_{A=A_{\rm cl}}=0$. Accordingly, the vector potential can be formally split as ${\bf A}(q)={\bf A}_\text{cl}({\bf q},0) + {\bf A}_\text{fl}({\bf q},\Omega_m)$, where ${\bf A}_\text{cl}$ represents the static classical background, and ${\bf A}_\text{fl}$ contains the fluctuations around this background.

The difference between thermal and quantum fluctuations appears in the equal-time propagator 
\begin{gather}
\mathcal{D}({\bf q})=T\sum_{\Omega_m} \mathcal{D}(q) = \frac{\coth\left( \frac{\sqrt{\Omega^2({\bf q})+4\pi Q_\text{M}({\bf q},T)}}{2T} \right)}{2\sqrt{\Omega^2({\bf q})+4\pi Q_\text{M}({\bf q},T)}}.
\end{gather}
The Meissner gap is temperature dependent and vanishes at the transition point, so that $Q_\text{M}({\bf q},T_c)=0$, while the cavity-induced gap is fixed $\Omega_0\leqslant \Omega({\bf q})$. In the low-temperature regime $T\sim T_c\ll \Omega_0$ (which we assume throughout the work) fluctuations are dominated by the zero-point motion, $\mathcal{D}({\bf q})\approx 1/2\Omega({\bf q})$, with the thermal contribution exponentially suppressed. Note that the GL formalism can include only low-momentum photons, so in this regime the kernel is reduced to the square of the London magnetic penetration depth $4\pi Q_\text{M}(0,T) = \lambda^{-2}_L$ [see Fig. (\ref{fig_1})].

\

\setcounter{equation}{0}
\renewcommand{\theequation}{B\arabic{equation}}

\textit{Appendix B: Effective electron-photon interaction}.-- 
Using the quantization of the fluctuating field ${\bf A}_\text{f{}l}$ in the action (\ref{S_tot}) one can write the photon energy as
\begin{gather}\label{S_ph_D}
    S_\text{ph}[\mathcal{A}] = \frac{1}{2T} \sum_{q,s,s'} \mathcal{A}_s(-q)\mathcal{D}^{-1}_{s,s'}(q)\mathcal{A}_{s'}(q),
\end{gather}
where the bare photon propagator reads as
\begin{gather}
    \mathcal{D}_{s,s'}({\bf q},i\Omega_m) = \langle \mathcal{A}_s(-q) \mathcal{A}_{s'}(q) \rangle = \frac{4\pi \delta_{ss'}}{\Omega_m^2 + \Omega^2({\bf q})}
\end{gather}
and  contains photon dispersion $\Omega^2({\bf q}) = \Omega_0^2+|{\bf q}|^2$ with the fundamental cavity frequency ${\Omega_0=\pi/L_z}$.

We neglect the diamagnetic term and focus on paramagnetic ${\bf j}\cdot {\bf A}_\text{fl}$ part of the electron-photon coupling, and within dipole approximation $|\nabla_{||} A({\bf r}_{||})| \ll k_F|A|$ using Fourier transform it can be written as  
\begin{gather}
    S_\text{pm}[\psi,\mathcal{A}] 
    \\ \notag
    = \frac{1}{\sqrt T}\sum_{q,k} \sum_{s,\sigma} g_s({\bf q}, {\bf k}-{\bf q}/2) \mathcal{A}_s({\bf q}) \bar\psi_{n-m,\sigma}({\bf k-q}) \psi_{n,\sigma}({\bf k})  ,
\end{gather}
where ${\bf j}$ is electron current and we introduced the electron-photon interaction vertex ${g_s({\bf q}, {\bf k}-{\bf q}/2) = - \sqrt{\frac2V}\frac{e}{m}{\bf e}_s({\bf q})\cdot ({\bf k}-{\bf q}/2)}$.

The total action is Gaussian in $\mathcal{A}_s$, therefore we can integrate out the fluctuating photonic degrees of freedom and obtain the effective electron action:
\begin{gather}
    S_\text{el-pht} = \sum
    _{\substack{k,k',q \\ \sigma,\sigma'}}\frac{V_{{\bf k},{\bf k}'}(q)}{2}
    \bar\psi_{k- q,\sigma}\psi_{{k},\sigma}\bar\psi_{{k}'+q,\sigma'}\psi_{{k}',\sigma'}
\end{gather}
where the interaction potential reads as ${V_{{\bf k},{\bf k}'}(q) = \sum_{s,s'} D_{s,s'}(q) g_s({\bf q}, {\bf k}-{\bf q}/2) g_{s'}(-{\bf q}, {\bf k}'+{\bf q}/2)}$. Since the energy transfer between electrons is much smaller than the photonic gap $\Omega_0$, one can neglect retardation effects and use the static limit $\Omega_m\rightarrow 0$ and thus get the standard Amperean coupling
\begin{gather}
    V_{{\bf k},{\bf k}'}({\bf q}) 
    = -V_0 \sum_s \frac{\Big( {\bf e}_{{\bf q},s} ({\bf k}-\frac{\bf q}{2}) \Big)\Big( {\bf e}_{-{\bf q},s} ({\bf k}'+\frac{\bf q}{2}) \Big)}{\Omega^2({\bf q})},
\end{gather}
which can be simplified to the Eq. (\ref{V_dyn}).

\

\setcounter{equation}{0}
\renewcommand{\theequation}{C\arabic{equation}}

\textit{Appendix C: Approximation for the self-energy}.--
First we perform the summation over Matsubara frequencies as $T\sum_n G_0({\bf k}, i\omega_n)=n_F(k)$, where $n_F$ is Fermi-Dirac distribution function. The interaction vertex $V_{{\bf k, k'}}({\bf q})$ reads as 
\begin{gather}
V_{{\bf k, k'}}({\bf k}-{\bf k}') = V_0 \frac{({\bf k}+{\bf k}')^2}{4\Omega^2({\bf k}-{\bf k}')},
\end{gather}
where the amplitude of the interaction vertex is
$V_0=\frac{8\pi e^2}{V m^2}$
and the volume of the cavity is $V=L^2\times L_z$. Performing Matsubara summation first and using continuum limit $\sum_{\bf k} \rightarrow \frac{L^2}{(2\pi)^2}\int d{\bf k}$ we end up with a 2D integral:
\begin{gather}\label{App_S_num}
    \Sigma({\bf k}) =  \frac{L^2}{(2\pi)^2}\int d{\bf k}' V_{{\bf k, k'}}({\bf k}-{\bf k}') n_F(k').
\end{gather}

For $T/E_F\lesssim T_c/E_F \ll 1$ and $T/E_F \lesssim \Omega_0/k_F$ one can use the Sommerfeld expansion for $n_F(k)$ and obtain a closed form of $\Sigma({\bf k})$.
More relevant case of $\Omega_0/k_F \ll T/E_F \ll 1$ does not provide a closed form of the integral, but its leading regularized part reads as 
\begin{gather}\label{App_S_2}
    \Sigma(k) \approx   32\pi \frac{\alpha E_F^2}{k_F^2L_z} ~\ln(k_F^2\Lambda^2) n_F(k),
\end{gather}
where $\Lambda$ is the ultraviolet truncation and in our case we have $\Lambda \sim 1/\Omega_0$. This result coincides with the $\delta$-function approximation, corresponding to a perfect forward scattering and used in \cite{PhysRevLett.122.133602}. According to the latter, the integrand in Eq. (\ref{App_S_num}) is peaked around ${|{\bf k}'-{\bf k}|\ll\Omega_0}$, thus one can employ the following approximation $\Omega^{-2}({\bf k}-{\bf k}')\approx c_0 \delta({\bf k}-{\bf k}')$, where the delta function is weighted by $c_0=\pi\ln\left( 1 + k_F^2/\Omega_0^2 \right)$. If the cavity momentum transfer is small, e.g. $|k-k_F|\ll k_F$, this delta function gives the main contribution to the integral and we get
\begin{gather}\label{AppA_approx}
    \Sigma({\bf k}) \approx  32 c_0 \frac{\alpha E_F^2}{k^2_F L_z}\left(\frac{{\bf k}^2}{k_F^2}\right)n_F(k).
\end{gather}

For the case of quasi-2D systems, such as weakly coupled layered structures with anisotropic mass tensor $m_z\gg m_{||}$ and almost cylindrical Fermi surface, we should extend the integration as following:
\begin{gather}\label{App_S_q2D}
    \Sigma({\bf k}_{||}) = \frac{L^2d_\text{S}}{(2\pi)^3}\int d k_z' d{\bf k}_{||}'  V_{{\bf k}_{||}, {\bf k}_{||}'}({\bf k}_{||}-{\bf k}_{||}') n_F({\bf k}),
\end{gather}
where ${\bf k}^2={\bf k}_{||}^2+{\bf k}_z^2$ and $d_\text{S}$ is the thickness on the superconductors. Note that $L^2$ is the lateral size of both the cavity and the material inside. This procedure will preserve the logarithmic divergence coming form the cylindrical shell ${\bf k}_{||}\approx {\bf k}_{||}'$, but effectively smear the distribution function $n_F(k)$. As a result we yield modified self-energy
\begin{gather}
    \Sigma({\bf k}_{||}) \approx  -32 c_0 \frac{\alpha E_F^2}{k_F^2L_z}\frac{{\bf k}^2}{k_F^2}\frac{k_F d_\text{S}}{2\pi}
   \sqrt{\frac{\pi T}{E_F}}~ \text{Li}_\frac12 \left[ -e^{ \left(1-\frac{k^2}{k_F^2}\right)\frac{E_F}{T}} \right],
\end{gather}
where we gain a prefactor proportional to $k_F d_\text{S}$, but at the same time the dependence on the momentum close to $k_F$ becomes less steep. Nevertheless, the overall effect can still potentially lead to an enhanced renormalization of the GL stiffness.

\

\setcounter{equation}{0}
\renewcommand{\theequation}{D\arabic{equation}}

\textit{Appendix D: Hubbard-Stratonovich transformation}.--
Applying the mean-field approximation to the electron part of the action \eqref{S_tot}, renormalized by the self-energy $\Sigma({\bf k})$, together with the BCS term from Eq. (\ref{S_bcs}) we obtain
\begin{gather}
    \tilde{S}_\text{el}[\psi]+S_\text{BCS}[\psi] \rightarrow \sum_{\bf p}  \frac{|\Delta({\bf p})|^2}{\lambda } 
    - \sum_{k} \hat{\bar\Psi}_{\bf k}\hat{\mathcal{G}}_0^{-1} \hat\Psi_{\bf k} ,
\end{gather}
where 
$\hat\Psi_{\bf k}({\bf p}) = ( \psi_{{\bf k+p}/2\uparrow} ~ \bar\psi_{{\bf -k+p}/2\downarrow} )^T $ and 
\begin{gather}\label{G_NG}
    \hat{{\mathcal{G}}}_0^{-1} = 
    \begin{pmatrix}
        \tilde G^{-1}_0({\bf k_+},i\omega_n) && \Delta({\bf p}) \\
        \bar\Delta({\bf p}) && -\tilde G^{-1}_0({\bf -k_-},-i\omega_n)
    \end{pmatrix}
\end{gather} 
are vector and matrix in Nambu space with ${\bf k}_{\pm}={\bf k}\pm{\bf p}/2$. The electron-photon interaction is encoded into particle propagator $\tilde G^{-1}_0$. By integrating out the fermionic degrees of freedom we obtain the effective action for the gap:
\begin{gather}\label{S_eff_Delta}
    S_\text{eff}[\Delta] = \sum_{\bf p}  \frac{|\Delta({\bf p})|^2}{\lambda } - \sum_{k} \text{tr}\ln\hat{\mathcal{G}}_0^{-1}  
    \\ \notag 
    \overset{T\rightarrow T_c}{=}  \sum_{\bf p} \bar\Delta({\bf p}) \big[ \mathcal{D}_\Delta({\bf p}) \big]^{-1} \Delta({\bf p}) + \mathcal{O}(\Delta^4),
\end{gather}
where we formally defined the superconducting gap propagator $\mathcal{D}_\Delta({\bf p}) = \big\langle \Delta({\bf p})\bar\Delta({\bf p}) \big\rangle$.

\

\setcounter{equation}{0}
\renewcommand{\theequation}{E\arabic{equation}}

\textit{Appendix E: Calculation of the polarization function}.--
For the vertex function $\Gamma({\bf k},{\bf p})$ we need to solve the Bethe-Salpeter equation (\ref{BSeq}). Corresponding interaction potential $V_{{\bf k}_+,-{\bf k}_-}({\bf k}-{\bf k}')$ can be simplified in the same way as the self-energy in Appendix C, and we can again utilize the delta-function approximation $\Omega^{-2}({\bf k}-{\bf k}')\approx c_0 \delta({\bf k}-{\bf k}')$. The latter transforms the integral equation for $\Gamma({\bf k},{\bf p})$ into an algebraic equation $\Gamma({\bf k},{\bf p}) = 1 - \mathcal{F}({\bf k},{\bf p})  \Gamma({\bf k},{\bf p})$, where
\begin{gather}\label{mathF}
    \mathcal{F}({\bf k},{\bf p}) = \Sigma_0 
    \left( \frac{{\bf k}_+{\bf k}_-}{k_F^2} \right)
    \frac{1-n_F(\tilde{\xi}_{{\bf k}_+})-n_F(\tilde{\xi}_{-{\bf k}_-})}{\tilde{\xi}_{{\bf k}_+}+\tilde{\xi}_{-{\bf k}_-}}.
\end{gather}
Here the latter part comes form the Matsubara summation $T\sum_n \tilde G_0({\bf k}_+,i\omega_n)\tilde G_0({\bf -k}_-,-i\omega_n)$ with the dressed electron dispersion $\tilde{\xi}_{\bf k}=\xi_{\bf k} - \Sigma({\bf k})$ and we used the amplitude of the self-energy from Eq. (\ref{AppA_approx}).

With the analytical expression for the vertex function we gets the polarization loop:
\begin{gather}\label{for_num}
    \tilde{\Pi}({\bf p}) 
    = \sum_{\bf k} \frac{1-n_F(\tilde{\xi}_{{\bf k}_+})-n_F(\tilde{\xi}_{-{\bf k}_-})}{\tilde{\xi}_{{\bf k}_+}+\tilde{\xi}_{-{\bf k}_-}} ~
    \frac{1}{1+ \mathcal{F}({\bf k},{\bf p})}.
\end{gather}
For $|{\bf p}|\ll \xi(0) \ll k_F$ we use $\xi_{\bf k\pm}\approx \xi_k\pm \pi T_c \xi(0){\bf n}_F{\bf p}$, where $\xi(0)$ is coherence length and ${\bf n}_F$ is the unit vector at the Fermi surface. Going to the integration over energy, we get $\sum_{\bf k} \rightarrow \nu_{2D} \int_{-\infty}^{\infty} d\xi_k \int_0^{2\pi}  \frac{d\theta}{2\pi}$, where $\nu_{2D} = m/2\pi$ is two-dimensional density of states. The ${\bf p}$-expansion can be formally written as 
\begin{gather}\label{for_num2}
    \tilde{\Pi}({\bf p}) \approx \tilde\Pi(0) - \tilde{a}_2(\Sigma_0){\bf p}^2 + \mathcal{O}({\bf p}^4).
\end{gather}
We replace $\tilde{\Pi}(0)$ with its bare value $\Pi(0)=\sum_{k}  G_0({\bf k_+},i\omega_n) G_0({\bf -k_-},-i\omega_n) \approx  \nu_{2D}\int_0^{\omega_D/2T} \text{th}(x)/x dx$ since we do not consider a shift of the critical temperature $T_c$. The temperature in $\tilde{\Pi}({\bf p})$ is set to $T = T_c$. Thus, we restore the Eq. (\ref{F_0}).


\begin{thebibliography}{45}%
\makeatletter
\providecommand \@ifxundefined [1]{%
 \@ifx{#1\undefined}
}%
\providecommand \@ifnum [1]{%
 \ifnum #1\expandafter \@firstoftwo
 \else \expandafter \@secondoftwo
 \fi
}%
\providecommand \@ifx [1]{%
 \ifx #1\expandafter \@firstoftwo
 \else \expandafter \@secondoftwo
 \fi
}%
\providecommand \natexlab [1]{#1}%
\providecommand \enquote  [1]{``#1''}%
\providecommand \bibnamefont  [1]{#1}%
\providecommand \bibfnamefont [1]{#1}%
\providecommand \citenamefont [1]{#1}%
\providecommand \href@noop [0]{\@secondoftwo}%
\providecommand \href [0]{\begingroup \@sanitize@url \@href}%
\providecommand \@href[1]{\@@startlink{#1}\@@href}%
\providecommand \@@href[1]{\endgroup#1\@@endlink}%
\providecommand \@sanitize@url [0]{\catcode `\\12\catcode `\$12\catcode `\&12\catcode `\#12\catcode `\^12\catcode `\_12\catcode `\%12\relax}%
\providecommand \@@startlink[1]{}%
\providecommand \@@endlink[0]{}%
\providecommand \url  [0]{\begingroup\@sanitize@url \@url }%
\providecommand \@url [1]{\endgroup\@href {#1}{\urlprefix }}%
\providecommand \urlprefix  [0]{URL }%
\providecommand \Eprint [0]{\href }%
\providecommand \doibase [0]{http://dx.doi.org/}%
\providecommand \selectlanguage [0]{\@gobble}%
\providecommand \bibinfo  [0]{\@secondoftwo}%
\providecommand \bibfield  [0]{\@secondoftwo}%
\providecommand \translation [1]{[#1]}%
\providecommand \BibitemOpen [0]{}%
\providecommand \bibitemStop [0]{}%
\providecommand \bibitemNoStop [0]{.\EOS\space}%
\providecommand \EOS [0]{\spacefactor3000\relax}%
\providecommand \BibitemShut  [1]{\csname bibitem#1\endcsname}%
\let\auto@bib@innerbib\@empty
\bibitem [{\citenamefont {Frisk~Kockum}\ \emph {et~al.}(2019)\citenamefont {Frisk~Kockum}, \citenamefont {Miranowicz}, \citenamefont {De~Liberato}, \citenamefont {Savasta},\ and\ \citenamefont {Nori}}]{frisk2019ultrastrong}%
  \BibitemOpen
  \bibfield  {author} {\bibinfo {author} {\bibfnamefont {A.}~\bibnamefont {Frisk~Kockum}}, \bibinfo {author} {\bibfnamefont {A.}~\bibnamefont {Miranowicz}}, \bibinfo {author} {\bibfnamefont {S.}~\bibnamefont {De~Liberato}}, \bibinfo {author} {\bibfnamefont {S.}~\bibnamefont {Savasta}}, \ and\ \bibinfo {author} {\bibfnamefont {F.}~\bibnamefont {Nori}},\ }\bibfield  {title} {\bibinfo {title} {Ultrastrong coupling between light and matter},\ }\href {https://www.nature.com/articles/s42254-018-0006-2} {\bibfield  {journal} {\bibinfo  {journal} {Nature Reviews Physics}\ }\textbf {\bibinfo {volume} {1}},\ \bibinfo {pages} {19} (\bibinfo {year} {2019})}\BibitemShut {NoStop}%
\bibitem [{\citenamefont {Garcia-Vidal}\ \emph {et~al.}(2021)\citenamefont {Garcia-Vidal}, \citenamefont {Ciuti},\ and\ \citenamefont {Ebbesen}}]{Garcia-Vidal:2021}%
  \BibitemOpen
  \bibfield  {author} {\bibinfo {author} {\bibfnamefont {F.~J.}\ \bibnamefont {Garcia-Vidal}}, \bibinfo {author} {\bibfnamefont {C.}~\bibnamefont {Ciuti}}, \ and\ \bibinfo {author} {\bibfnamefont {T.~W.}\ \bibnamefont {Ebbesen}},\ }\bibfield  {title} {\bibinfo {title} {Manipulating matter by strong coupling to vacuum fields},\ }\href {\doibase 10.1126/science.abd0336} {\bibfield  {journal} {\bibinfo  {journal} {Science}\ }\textbf {\bibinfo {volume} {373}},\ \bibinfo {pages} {eabd0336} (\bibinfo {year} {2021})}\BibitemShut {NoStop}%
\bibitem [{\citenamefont {Mivehvar}\ \emph {et~al.}(2021)\citenamefont {Mivehvar}, \citenamefont {Piazza}, \citenamefont {Donner},\ and\ \citenamefont {Ritsch}}]{Mivehvar:2021}%
  \BibitemOpen
  \bibfield  {author} {\bibinfo {author} {\bibfnamefont {F.}~\bibnamefont {Mivehvar}}, \bibinfo {author} {\bibfnamefont {F.}~\bibnamefont {Piazza}}, \bibinfo {author} {\bibfnamefont {T.}~\bibnamefont {Donner}}, \ and\ \bibinfo {author} {\bibfnamefont {H.}~\bibnamefont {Ritsch}},\ }\bibfield  {title} {\bibinfo {title} {Cavity {QED} with quantum gases: new paradigms in many-body physics},\ }\href {\doibase 10.1080/00018732.2021.1969727} {\bibfield  {journal} {\bibinfo  {journal} {Advances in Physics}\ }\textbf {\bibinfo {volume} {70}},\ \bibinfo {pages} {1} (\bibinfo {year} {2021})}\BibitemShut {NoStop}%
\bibitem [{\citenamefont {Schlawin}\ \emph {et~al.}(2022)\citenamefont {Schlawin}, \citenamefont {Kennes},\ and\ \citenamefont {Sentef}}]{Schlawin:2022}%
  \BibitemOpen
  \bibfield  {author} {\bibinfo {author} {\bibfnamefont {F.}~\bibnamefont {Schlawin}}, \bibinfo {author} {\bibfnamefont {D.~M.}\ \bibnamefont {Kennes}}, \ and\ \bibinfo {author} {\bibfnamefont {M.~A.}\ \bibnamefont {Sentef}},\ }\bibfield  {title} {\bibinfo {title} {Cavity quantum materials},\ }\href {\doibase 10.1063/5.0078326} {\bibfield  {journal} {\bibinfo  {journal} {Applied Physics Reviews}\ }\textbf {\bibinfo {volume} {9}},\ \bibinfo {pages} {011312} (\bibinfo {year} {2022})}\BibitemShut {NoStop}%
\bibitem [{\citenamefont {Bloch}\ \emph {et~al.}(2022)\citenamefont {Bloch}, \citenamefont {Cavalleri}, \citenamefont {Galitski}, \citenamefont {Hafezi},\ and\ \citenamefont {Rubio}}]{Bloch:2022}%
  \BibitemOpen
  \bibfield  {author} {\bibinfo {author} {\bibfnamefont {J.}~\bibnamefont {Bloch}}, \bibinfo {author} {\bibfnamefont {A.}~\bibnamefont {Cavalleri}}, \bibinfo {author} {\bibfnamefont {V.}~\bibnamefont {Galitski}}, \bibinfo {author} {\bibfnamefont {M.}~\bibnamefont {Hafezi}}, \ and\ \bibinfo {author} {\bibfnamefont {A.}~\bibnamefont {Rubio}},\ }\bibfield  {title} {\bibinfo {title} {Strongly correlated electron-photon systems},\ }\href {\doibase 10.1038/s41586-022-04639-6} {\bibfield  {journal} {\bibinfo  {journal} {Nature}\ }\textbf {\bibinfo {volume} {606}},\ \bibinfo {pages} {41} (\bibinfo {year} {2022})}\BibitemShut {NoStop}%
\bibitem [{\citenamefont {Appugliese}\ \emph {et~al.}(2022)\citenamefont {Appugliese}, \citenamefont {Enkner}, \citenamefont {Paravicini-Bagliani}, \citenamefont {Beck}, \citenamefont {Reichl}, \citenamefont {Wegscheider}, \citenamefont {Scalari}, \citenamefont {Ciuti},\ and\ \citenamefont {Faist}}]{appugliese2022breakdown}%
  \BibitemOpen
  \bibfield  {author} {\bibinfo {author} {\bibfnamefont {F.}~\bibnamefont {Appugliese}}, \bibinfo {author} {\bibfnamefont {J.}~\bibnamefont {Enkner}}, \bibinfo {author} {\bibfnamefont {G.~L.}\ \bibnamefont {Paravicini-Bagliani}}, \bibinfo {author} {\bibfnamefont {M.}~\bibnamefont {Beck}}, \bibinfo {author} {\bibfnamefont {C.}~\bibnamefont {Reichl}}, \bibinfo {author} {\bibfnamefont {W.}~\bibnamefont {Wegscheider}}, \bibinfo {author} {\bibfnamefont {G.}~\bibnamefont {Scalari}}, \bibinfo {author} {\bibfnamefont {C.}~\bibnamefont {Ciuti}}, \ and\ \bibinfo {author} {\bibfnamefont {J.}~\bibnamefont {Faist}},\ }\bibfield  {title} {\bibinfo {title} {Breakdown of topological protection by cavity vacuum fields in the integer quantum hall effect},\ }\href {https://www.science.org/doi/10.1126/science.abl5818} {\bibfield  {journal} {\bibinfo  {journal} {Science}\ }\textbf {\bibinfo {volume} {375}},\ \bibinfo {pages} {1030} (\bibinfo {year} {2022})}\BibitemShut {NoStop}%
\bibitem [{\citenamefont {Thomas}\ \emph {et~al.}(2021)\citenamefont {Thomas}, \citenamefont {Devaux}, \citenamefont {Nagarajan}, \citenamefont {Rogez}, \citenamefont {Seidel}, \citenamefont {Richard}, \citenamefont {Genet}, \citenamefont {Drillon},\ and\ \citenamefont {Ebbesen}}]{Thomas:2021}%
  \BibitemOpen
  \bibfield  {author} {\bibinfo {author} {\bibfnamefont {A.}~\bibnamefont {Thomas}}, \bibinfo {author} {\bibfnamefont {E.}~\bibnamefont {Devaux}}, \bibinfo {author} {\bibfnamefont {K.}~\bibnamefont {Nagarajan}}, \bibinfo {author} {\bibfnamefont {G.}~\bibnamefont {Rogez}}, \bibinfo {author} {\bibfnamefont {M.}~\bibnamefont {Seidel}}, \bibinfo {author} {\bibfnamefont {F.}~\bibnamefont {Richard}}, \bibinfo {author} {\bibfnamefont {C.}~\bibnamefont {Genet}}, \bibinfo {author} {\bibfnamefont {M.}~\bibnamefont {Drillon}}, \ and\ \bibinfo {author} {\bibfnamefont {T.~W.}\ \bibnamefont {Ebbesen}},\ }\bibfield  {title} {\bibinfo {title} {Large enhancement of ferromagnetism under a collective strong coupling of {YBCO} nanoparticles},\ }\href {\doibase 10.1021/acs.nanolett.1c01083} {\bibfield  {journal} {\bibinfo  {journal} {Nano Letters}\ }\textbf {\bibinfo {volume} {21}},\ \bibinfo {pages} {4365} (\bibinfo {year} {2021})}\BibitemShut {NoStop}%
\bibitem [{\citenamefont {Jarc}\ \emph {et~al.}(2023)\citenamefont {Jarc}, \citenamefont {Mathengattil}, \citenamefont {Montanaro}, \citenamefont {Giusti}, \citenamefont {Rigoni}, \citenamefont {Sergo}, \citenamefont {Fassioli}, \citenamefont {Winnerl}, \citenamefont {Dal~Zilio}, \citenamefont {Mihailovic},\ and\ \citenamefont {Prelovsek}}]{Jarc:2023}%
  \BibitemOpen
  \bibfield  {author} {\bibinfo {author} {\bibfnamefont {G.}~\bibnamefont {Jarc}}, \bibinfo {author} {\bibfnamefont {S.~Y.}\ \bibnamefont {Mathengattil}}, \bibinfo {author} {\bibfnamefont {A.}~\bibnamefont {Montanaro}}, \bibinfo {author} {\bibfnamefont {F.}~\bibnamefont {Giusti}}, \bibinfo {author} {\bibfnamefont {E.~M.}\ \bibnamefont {Rigoni}}, \bibinfo {author} {\bibfnamefont {R.}~\bibnamefont {Sergo}}, \bibinfo {author} {\bibfnamefont {F.}~\bibnamefont {Fassioli}}, \bibinfo {author} {\bibfnamefont {S.}~\bibnamefont {Winnerl}}, \bibinfo {author} {\bibfnamefont {S.}~\bibnamefont {Dal~Zilio}}, \bibinfo {author} {\bibfnamefont {D.}~\bibnamefont {Mihailovic}}, \ and\ \bibinfo {author} {\bibfnamefont {P.}~\bibnamefont {Prelovsek}},\ }\bibfield  {title} {\bibinfo {title} {Cavity-mediated thermal control of metal-to-insulator transition in {1T-TaS}$_2$},\ }\href {\doibase 10.1038/s41586-023-06596-2} {\bibfield  {journal} {\bibinfo  {journal} {Nature}\ }\textbf {\bibinfo {volume} {622}},\ \bibinfo {pages} {487} (\bibinfo {year} {2023})}\BibitemShut {NoStop}%
\bibitem [{\citenamefont {Thomas}\ \emph {et~al.}(2025)\citenamefont {Thomas}, \citenamefont {Devaux}, \citenamefont {Nagarajan}, \citenamefont {Chervy}, \citenamefont {Seidel}, \citenamefont {Rogez}, \citenamefont {Robert}, \citenamefont {Drillon}, \citenamefont {Ruan}, \citenamefont {Schlittenhardt}, \citenamefont {Ruben}, \citenamefont {Hagenmüller}, \citenamefont {Schütz}, \citenamefont {Schachenmayer}, \citenamefont {Genet}, \citenamefont {Pupillo},\ and\ \citenamefont {Ebbesen}}]{Thomas:2019}%
  \BibitemOpen
  \bibfield  {author} {\bibinfo {author} {\bibfnamefont {A.}~\bibnamefont {Thomas}}, \bibinfo {author} {\bibfnamefont {E.}~\bibnamefont {Devaux}}, \bibinfo {author} {\bibfnamefont {K.}~\bibnamefont {Nagarajan}}, \bibinfo {author} {\bibfnamefont {T.}~\bibnamefont {Chervy}}, \bibinfo {author} {\bibfnamefont {M.}~\bibnamefont {Seidel}}, \bibinfo {author} {\bibfnamefont {G.}~\bibnamefont {Rogez}}, \bibinfo {author} {\bibfnamefont {J.}~\bibnamefont {Robert}}, \bibinfo {author} {\bibfnamefont {M.}~\bibnamefont {Drillon}}, \bibinfo {author} {\bibfnamefont {T.~T.}\ \bibnamefont {Ruan}}, \bibinfo {author} {\bibfnamefont {S.}~\bibnamefont {Schlittenhardt}}, \bibinfo {author} {\bibfnamefont {M.}~\bibnamefont {Ruben}}, \bibinfo {author} {\bibfnamefont {D.}~\bibnamefont {Hagenmüller}}, \bibinfo {author} {\bibfnamefont {S.}~\bibnamefont {Schütz}}, \bibinfo {author} {\bibfnamefont {J.}~\bibnamefont {Schachenmayer}}, \bibinfo {author} {\bibfnamefont {C.}~\bibnamefont {Genet}}, \bibinfo {author} {\bibfnamefont {G.}~\bibnamefont {Pupillo}}, \ and\ \bibinfo {author} {\bibfnamefont {T.~W.}\ \bibnamefont {Ebbesen}},\ }\bibfield  {title} {\bibinfo {title} {Exploring superconductivity under strong coupling with the vacuum electromagnetic field},\ }\href {\doibase 10.1063/5.0231202} {\bibfield  {journal} {\bibinfo  {journal} {The Journal of Chemical Physics}\ }\textbf {\bibinfo {volume} {162}},\ \bibinfo {pages} {134701} (\bibinfo {year} {2025})}\BibitemShut {NoStop}%
\bibitem [{\citenamefont {Keren}\ \emph {et~al.}(2026)\citenamefont {Keren}, \citenamefont {Webb}, \citenamefont {Zhang}, \citenamefont {Xu}, \citenamefont {Sun}, \citenamefont {Kim}, \citenamefont {Shin}, \citenamefont {Zhang}, \citenamefont {Zhang}, \citenamefont {Pereira}, \citenamefont {Yao}, \citenamefont {Okugawa}, \citenamefont {Michael}, \citenamefont {Vi{\~{n}}as~Bostr{\"o}m}, \citenamefont {Edgar}, \citenamefont {Wolf}, \citenamefont {Julian}, \citenamefont {Prasankumar}, \citenamefont {Miyagawa}, \citenamefont {Kanoda}, \citenamefont {Gu}, \citenamefont {Cothrine}, \citenamefont {Mandrus}, \citenamefont {Buzzi}, \citenamefont {Cavalleri}, \citenamefont {Dean}, \citenamefont {Kennes}, \citenamefont {Millis}, \citenamefont {Li}, \citenamefont {Sentef}, \citenamefont {Rubio}, \citenamefont {Pasupathy},\ and\ \citenamefont {Basov}}]{Keren2026}%
  \BibitemOpen
  \bibfield  {author} {\bibinfo {author} {\bibfnamefont {I.}~\bibnamefont {Keren}}, \bibinfo {author} {\bibfnamefont {T.~A.}\ \bibnamefont {Webb}}, \bibinfo {author} {\bibfnamefont {S.}~\bibnamefont {Zhang}}, \bibinfo {author} {\bibfnamefont {J.}~\bibnamefont {Xu}}, \bibinfo {author} {\bibfnamefont {D.}~\bibnamefont {Sun}}, \bibinfo {author} {\bibfnamefont {B.~S.~Y.}\ \bibnamefont {Kim}}, \bibinfo {author} {\bibfnamefont {D.}~\bibnamefont {Shin}}, \bibinfo {author} {\bibfnamefont {S.~S.}\ \bibnamefont {Zhang}}, \bibinfo {author} {\bibfnamefont {J.}~\bibnamefont {Zhang}}, \bibinfo {author} {\bibfnamefont {G.}~\bibnamefont {Pereira}}, \bibinfo {author} {\bibfnamefont {J.}~\bibnamefont {Yao}}, \bibinfo {author} {\bibfnamefont {T.}~\bibnamefont {Okugawa}}, \bibinfo {author} {\bibfnamefont {M.~H.}\ \bibnamefont {Michael}}, \bibinfo {author} {\bibfnamefont {E.}~\bibnamefont {Vi{\~{n}}as~Bostr{\"o}m}}, \bibinfo {author} {\bibfnamefont {J.~H.}\ \bibnamefont {Edgar}}, \bibinfo {author} {\bibfnamefont {S.}~\bibnamefont {Wolf}}, \bibinfo {author} {\bibfnamefont {M.}~\bibnamefont {Julian}}, \bibinfo {author} {\bibfnamefont {R.~P.}\ \bibnamefont {Prasankumar}}, \bibinfo {author} {\bibfnamefont {K.}~\bibnamefont {Miyagawa}}, \bibinfo {author} {\bibfnamefont {K.}~\bibnamefont {Kanoda}}, \bibinfo {author} {\bibfnamefont {G.}~\bibnamefont {Gu}}, \bibinfo {author} {\bibfnamefont {M.}~\bibnamefont {Cothrine}}, \bibinfo {author} {\bibfnamefont {D.}~\bibnamefont {Mandrus}}, \bibinfo {author} {\bibfnamefont {M.}~\bibnamefont {Buzzi}}, \bibinfo {author} {\bibfnamefont {A.}~\bibnamefont {Cavalleri}}, \bibinfo {author} {\bibfnamefont {C.~R.}\ \bibnamefont {Dean}}, \bibinfo {author} {\bibfnamefont {D.~M.}\ \bibnamefont {Kennes}}, \bibinfo {author} {\bibfnamefont {A.~J.}\ \bibnamefont {Millis}}, \bibinfo {author} {\bibfnamefont {Q.}~\bibnamefont {Li}}, \bibinfo {author} {\bibfnamefont {M.~A.}\ \bibnamefont {Sentef}}, \bibinfo {author} {\bibfnamefont {A.}~\bibnamefont {Rubio}}, \bibinfo {author} {\bibfnamefont {A.~N.}\ \bibnamefont {Pasupathy}}, \ and\ \bibinfo {author} {\bibfnamefont {D.~N.}\ \bibnamefont {Basov}},\ }\bibfield  {title} {\bibinfo {title} {Cavity-altered superconductivity},\ }\href {\doibase 10.1038/s41586-025-10062-6} {\bibfield  {journal} {\bibinfo  {journal} {Nature}\ }\textbf {\bibinfo {volume} {650}},\ \bibinfo {pages} {864} (\bibinfo {year} {2026})}\BibitemShut {NoStop}%
\bibitem [{\citenamefont {Xu}\ \emph {et~al.}(2026)\citenamefont {Xu}, \citenamefont {Baydin}, \citenamefont {Yi}, \citenamefont {Lu}, \citenamefont {Zhu}, \citenamefont {Kritzell}, \citenamefont {Doumani}, \citenamefont {Kim}, \citenamefont {Tay}, \citenamefont {Rubio},\ and\ \citenamefont {Kono}}]{xu2026}%
  \BibitemOpen
  \bibfield  {author} {\bibinfo {author} {\bibfnamefont {H.}~\bibnamefont {Xu}}, \bibinfo {author} {\bibfnamefont {A.}~\bibnamefont {Baydin}}, \bibinfo {author} {\bibfnamefont {Q.}~\bibnamefont {Yi}}, \bibinfo {author} {\bibfnamefont {I.-T.}\ \bibnamefont {Lu}}, \bibinfo {author} {\bibfnamefont {N.}~\bibnamefont {Zhu}}, \bibinfo {author} {\bibfnamefont {T.~E.}\ \bibnamefont {Kritzell}}, \bibinfo {author} {\bibfnamefont {J.}~\bibnamefont {Doumani}}, \bibinfo {author} {\bibfnamefont {D.}~\bibnamefont {Kim}}, \bibinfo {author} {\bibfnamefont {F.}~\bibnamefont {Tay}}, \bibinfo {author} {\bibfnamefont {A.}~\bibnamefont {Rubio}}, \ and\ \bibinfo {author} {\bibfnamefont {J.}~\bibnamefont {Kono}},\ }\href@noop {} {\bibinfo {title} {{Vacuum-dressed superconductivity in NbN observed in a high-$Q$ terahertz cavity}},\ } (\bibinfo {year} {2026}),\ \Eprint {http://arxiv.org/abs/2601.08191} {arXiv:2601.08191} \BibitemShut {NoStop}%
\bibitem [{\citenamefont {Sentef}\ \emph {et~al.}(2018)\citenamefont {Sentef}, \citenamefont {Ruggenthaler},\ and\ \citenamefont {Rubio}}]{sentef2018cavity}%
  \BibitemOpen
  \bibfield  {author} {\bibinfo {author} {\bibfnamefont {M.~A.}\ \bibnamefont {Sentef}}, \bibinfo {author} {\bibfnamefont {M.}~\bibnamefont {Ruggenthaler}}, \ and\ \bibinfo {author} {\bibfnamefont {A.}~\bibnamefont {Rubio}},\ }\bibfield  {title} {\bibinfo {title} {Cavity quantum-electrodynamical polaritonically enhanced electron-phonon coupling and its influence on superconductivity},\ }\href {https://www.science.org/doi/10.1126/sciadv.aau6969} {\bibfield  {journal} {\bibinfo  {journal} {Science advances}\ }\textbf {\bibinfo {volume} {4}},\ \bibinfo {pages} {eaau6969} (\bibinfo {year} {2018})}\BibitemShut {NoStop}%
\bibitem [{\citenamefont {Schlawin}\ \emph {et~al.}(2019)\citenamefont {Schlawin}, \citenamefont {Cavalleri},\ and\ \citenamefont {Jaksch}}]{PhysRevLett.122.133602}%
  \BibitemOpen
  \bibfield  {author} {\bibinfo {author} {\bibfnamefont {F.}~\bibnamefont {Schlawin}}, \bibinfo {author} {\bibfnamefont {A.}~\bibnamefont {Cavalleri}}, \ and\ \bibinfo {author} {\bibfnamefont {D.}~\bibnamefont {Jaksch}},\ }\bibfield  {title} {\bibinfo {title} {Cavity-mediated electron-photon superconductivity},\ }\href {\doibase 10.1103/PhysRevLett.122.133602} {\bibfield  {journal} {\bibinfo  {journal} {Phys. Rev. Lett.}\ }\textbf {\bibinfo {volume} {122}},\ \bibinfo {pages} {133602} (\bibinfo {year} {2019})}\BibitemShut {NoStop}%
\bibitem [{\citenamefont {Chakraborty}\ and\ \citenamefont {Piazza}(2021)}]{chakraborty2021long}%
  \BibitemOpen
  \bibfield  {author} {\bibinfo {author} {\bibfnamefont {A.}~\bibnamefont {Chakraborty}}\ and\ \bibinfo {author} {\bibfnamefont {F.}~\bibnamefont {Piazza}},\ }\bibfield  {title} {\bibinfo {title} {Long-range photon fluctuations enhance photon-mediated electron pairing and superconductivity},\ }\href {\doibase 10.1103/PhysRevLett.127.177002} {\bibfield  {journal} {\bibinfo  {journal} {Phys. Rev. Lett.}\ }\textbf {\bibinfo {volume} {127}},\ \bibinfo {pages} {177002} (\bibinfo {year} {2021})}\BibitemShut {NoStop}%
\bibitem [{\citenamefont {Andolina}\ \emph {et~al.}(2024)\citenamefont {Andolina}, \citenamefont {De~Pasquale}, \citenamefont {Pellegrino}, \citenamefont {Torre}, \citenamefont {Koppens},\ and\ \citenamefont {Polini}}]{PhysRevB.109.104513}%
  \BibitemOpen
  \bibfield  {author} {\bibinfo {author} {\bibfnamefont {G.~M.}\ \bibnamefont {Andolina}}, \bibinfo {author} {\bibfnamefont {A.}~\bibnamefont {De~Pasquale}}, \bibinfo {author} {\bibfnamefont {F.~M.~D.}\ \bibnamefont {Pellegrino}}, \bibinfo {author} {\bibfnamefont {I.}~\bibnamefont {Torre}}, \bibinfo {author} {\bibfnamefont {F.~H.~L.}\ \bibnamefont {Koppens}}, \ and\ \bibinfo {author} {\bibfnamefont {M.}~\bibnamefont {Polini}},\ }\bibfield  {title} {\bibinfo {title} {Amperean superconductivity cannot be induced by deep subwavelength cavities in a two-dimensional material},\ }\href {\doibase 10.1103/PhysRevB.109.104513} {\bibfield  {journal} {\bibinfo  {journal} {Phys. Rev. B}\ }\textbf {\bibinfo {volume} {109}},\ \bibinfo {pages} {104513} (\bibinfo {year} {2024})}\BibitemShut {NoStop}%
\bibitem [{\citenamefont {Eckhardt}\ \emph {et~al.}(2024)\citenamefont {Eckhardt}, \citenamefont {Chattopadhyay}, \citenamefont {Kennes}, \citenamefont {Demler}, \citenamefont {Sentef},\ and\ \citenamefont {Michael}}]{eckhardt2024theory}%
  \BibitemOpen
  \bibfield  {author} {\bibinfo {author} {\bibfnamefont {C.~J.}\ \bibnamefont {Eckhardt}}, \bibinfo {author} {\bibfnamefont {S.}~\bibnamefont {Chattopadhyay}}, \bibinfo {author} {\bibfnamefont {D.~M.}\ \bibnamefont {Kennes}}, \bibinfo {author} {\bibfnamefont {E.~A.}\ \bibnamefont {Demler}}, \bibinfo {author} {\bibfnamefont {M.~A.}\ \bibnamefont {Sentef}}, \ and\ \bibinfo {author} {\bibfnamefont {M.~H.}\ \bibnamefont {Michael}},\ }\bibfield  {title} {\bibinfo {title} {Theory of resonantly enhanced photo-induced superconductivity},\ }\href {https://www.nature.com/articles/s41467-024-46632-x} {\bibfield  {journal} {\bibinfo  {journal} {Nature Communications}\ }\textbf {\bibinfo {volume} {15}},\ \bibinfo {pages} {2300} (\bibinfo {year} {2024})}\BibitemShut {NoStop}%
\bibitem [{\citenamefont {Lu}\ \emph {et~al.}(2024)\citenamefont {Lu}, \citenamefont {Shin}, \citenamefont {Svendsen}, \citenamefont {H{\"u}bener}, \citenamefont {De~Giovannini}, \citenamefont {Latini}, \citenamefont {Ruggenthaler},\ and\ \citenamefont {Rubio}}]{lu2024cavity}%
  \BibitemOpen
  \bibfield  {author} {\bibinfo {author} {\bibfnamefont {I.-T.}\ \bibnamefont {Lu}}, \bibinfo {author} {\bibfnamefont {D.}~\bibnamefont {Shin}}, \bibinfo {author} {\bibfnamefont {M.~K.}\ \bibnamefont {Svendsen}}, \bibinfo {author} {\bibfnamefont {H.}~\bibnamefont {H{\"u}bener}}, \bibinfo {author} {\bibfnamefont {U.}~\bibnamefont {De~Giovannini}}, \bibinfo {author} {\bibfnamefont {S.}~\bibnamefont {Latini}}, \bibinfo {author} {\bibfnamefont {M.}~\bibnamefont {Ruggenthaler}}, \ and\ \bibinfo {author} {\bibfnamefont {A.}~\bibnamefont {Rubio}},\ }\bibfield  {title} {\bibinfo {title} {Cavity-enhanced superconductivity in mgb2 from first-principles quantum electrodynamics (qedft)},\ }\href {https://www.pnas.org/doi/10.1073/pnas.2415061121} {\bibfield  {journal} {\bibinfo  {journal} {Proceedings of the National Academy of Sciences}\ }\textbf {\bibinfo {volume} {121}},\ \bibinfo {pages} {e2415061121} (\bibinfo {year} {2024})}\BibitemShut {NoStop}%
\bibitem [{\citenamefont {Kozin}\ \emph {et~al.}(2025)\citenamefont {Kozin}, \citenamefont {Thingstad}, \citenamefont {Loss},\ and\ \citenamefont {Klinovaja}}]{kozin2025cavity}%
  \BibitemOpen
  \bibfield  {author} {\bibinfo {author} {\bibfnamefont {V.~K.}\ \bibnamefont {Kozin}}, \bibinfo {author} {\bibfnamefont {E.}~\bibnamefont {Thingstad}}, \bibinfo {author} {\bibfnamefont {D.}~\bibnamefont {Loss}}, \ and\ \bibinfo {author} {\bibfnamefont {J.}~\bibnamefont {Klinovaja}},\ }\bibfield  {title} {\bibinfo {title} {Cavity-enhanced superconductivity via band engineering},\ }\href {https://link.aps.org/doi/10.1103/PhysRevB.111.035410} {\bibfield  {journal} {\bibinfo  {journal} {Phys. Rev. B}\ }\textbf {\bibinfo {volume} {111}},\ \bibinfo {pages} {035410} (\bibinfo {year} {2025})}\BibitemShut {NoStop}%
\bibitem [{\citenamefont {Laplace}\ \emph {et~al.}(2016)\citenamefont {Laplace}, \citenamefont {Fernandez-Pena}, \citenamefont {Gariglio}, \citenamefont {Triscone},\ and\ \citenamefont {Cavalleri}}]{PhysRevB.93.075152}%
  \BibitemOpen
  \bibfield  {author} {\bibinfo {author} {\bibfnamefont {Y.}~\bibnamefont {Laplace}}, \bibinfo {author} {\bibfnamefont {S.}~\bibnamefont {Fernandez-Pena}}, \bibinfo {author} {\bibfnamefont {S.}~\bibnamefont {Gariglio}}, \bibinfo {author} {\bibfnamefont {J.~M.}\ \bibnamefont {Triscone}}, \ and\ \bibinfo {author} {\bibfnamefont {A.}~\bibnamefont {Cavalleri}},\ }\bibfield  {title} {\bibinfo {title} {Proposed cavity josephson plasmonics with complex-oxide heterostructures},\ }\href {\doibase 10.1103/PhysRevB.93.075152} {\bibfield  {journal} {\bibinfo  {journal} {Phys. Rev. B}\ }\textbf {\bibinfo {volume} {93}},\ \bibinfo {pages} {075152} (\bibinfo {year} {2016})}\BibitemShut {NoStop}%
\bibitem [{\citenamefont {Allocca}\ \emph {et~al.}(2019)\citenamefont {Allocca}, \citenamefont {Raines}, \citenamefont {Curtis},\ and\ \citenamefont {Galitski}}]{PhysRevB.99.020504}%
  \BibitemOpen
  \bibfield  {author} {\bibinfo {author} {\bibfnamefont {A.~A.}\ \bibnamefont {Allocca}}, \bibinfo {author} {\bibfnamefont {Z.~M.}\ \bibnamefont {Raines}}, \bibinfo {author} {\bibfnamefont {J.~B.}\ \bibnamefont {Curtis}}, \ and\ \bibinfo {author} {\bibfnamefont {V.~M.}\ \bibnamefont {Galitski}},\ }\bibfield  {title} {\bibinfo {title} {Cavity superconductor-polaritons},\ }\href {\doibase 10.1103/PhysRevB.99.020504} {\bibfield  {journal} {\bibinfo  {journal} {Phys. Rev. B}\ }\textbf {\bibinfo {volume} {99}},\ \bibinfo {pages} {020504} (\bibinfo {year} {2019})}\BibitemShut {NoStop}%
\bibitem [{\citenamefont {Raines}\ \emph {et~al.}(2020)\citenamefont {Raines}, \citenamefont {Allocca}, \citenamefont {Hafezi},\ and\ \citenamefont {Galitski}}]{PhysRevResearch.2.013143}%
  \BibitemOpen
  \bibfield  {author} {\bibinfo {author} {\bibfnamefont {Z.~M.}\ \bibnamefont {Raines}}, \bibinfo {author} {\bibfnamefont {A.~A.}\ \bibnamefont {Allocca}}, \bibinfo {author} {\bibfnamefont {M.}~\bibnamefont {Hafezi}}, \ and\ \bibinfo {author} {\bibfnamefont {V.~M.}\ \bibnamefont {Galitski}},\ }\bibfield  {title} {\bibinfo {title} {Cavity higgs polaritons},\ }\href {\doibase 10.1103/PhysRevResearch.2.013143} {\bibfield  {journal} {\bibinfo  {journal} {Phys. Rev. Res.}\ }\textbf {\bibinfo {volume} {2}},\ \bibinfo {pages} {013143} (\bibinfo {year} {2020})}\BibitemShut {NoStop}%
\bibitem [{\citenamefont {Gao}\ \emph {et~al.}(2021)\citenamefont {Gao}, \citenamefont {Schlawin},\ and\ \citenamefont {Jaksch}}]{PhysRevB.104.L140503}%
  \BibitemOpen
  \bibfield  {author} {\bibinfo {author} {\bibfnamefont {H.}~\bibnamefont {Gao}}, \bibinfo {author} {\bibfnamefont {F.}~\bibnamefont {Schlawin}}, \ and\ \bibinfo {author} {\bibfnamefont {D.}~\bibnamefont {Jaksch}},\ }\bibfield  {title} {\bibinfo {title} {Higgs mode stabilization by photoinduced long-range interactions in a superconductor},\ }\href {\doibase 10.1103/PhysRevB.104.L140503} {\bibfield  {journal} {\bibinfo  {journal} {Phys. Rev. B}\ }\textbf {\bibinfo {volume} {104}},\ \bibinfo {pages} {L140503} (\bibinfo {year} {2021})}\BibitemShut {NoStop}%
\bibitem [{\citenamefont {Dmytruk}\ and\ \citenamefont {Schir{\`o}}(2024)}]{dmytruk2024hybrid}%
  \BibitemOpen
  \bibfield  {author} {\bibinfo {author} {\bibfnamefont {O.}~\bibnamefont {Dmytruk}}\ and\ \bibinfo {author} {\bibfnamefont {M.}~\bibnamefont {Schir{\`o}}},\ }\bibfield  {title} {\bibinfo {title} {Hybrid light-matter states in topological superconductors coupled to cavity photons},\ }\href {https://link.aps.org/doi/10.1103/PhysRevB.110.075416} {\bibfield  {journal} {\bibinfo  {journal} {Phys. Rev. B}\ }\textbf {\bibinfo {volume} {110}},\ \bibinfo {pages} {075416} (\bibinfo {year} {2024})}\BibitemShut {NoStop}%
\bibitem [{\citenamefont {Bacciconi}\ \emph {et~al.}(2024)\citenamefont {Bacciconi}, \citenamefont {Andolina},\ and\ \citenamefont {Mora}}]{bacciconi2024topological}%
  \BibitemOpen
  \bibfield  {author} {\bibinfo {author} {\bibfnamefont {Z.}~\bibnamefont {Bacciconi}}, \bibinfo {author} {\bibfnamefont {G.~M.}\ \bibnamefont {Andolina}}, \ and\ \bibinfo {author} {\bibfnamefont {C.}~\bibnamefont {Mora}},\ }\bibfield  {title} {\bibinfo {title} {Topological protection of majorana polaritons in a cavity},\ }\href {https://link.aps.org/doi/10.1103/PhysRevB.109.165434} {\bibfield  {journal} {\bibinfo  {journal} {Phys. Rev. B}\ }\textbf {\bibinfo {volume} {109}},\ \bibinfo {pages} {165434} (\bibinfo {year} {2024})}\BibitemShut {NoStop}%
\bibitem [{\citenamefont {Zwettler}\ \emph {et~al.}(2025)\citenamefont {Zwettler}, \citenamefont {Marijanović}, \citenamefont {Bühler}, \citenamefont {Chattopadhyay}, \citenamefont {Pace}, \citenamefont {Skolc}, \citenamefont {Helson}, \citenamefont {Uchino}, \citenamefont {Demler},\ and\ \citenamefont {Brantut}}]{ZwettlerBrantut2025}%
  \BibitemOpen
  \bibfield  {author} {\bibinfo {author} {\bibfnamefont {T.}~\bibnamefont {Zwettler}}, \bibinfo {author} {\bibfnamefont {F.}~\bibnamefont {Marijanović}}, \bibinfo {author} {\bibfnamefont {T.}~\bibnamefont {Bühler}}, \bibinfo {author} {\bibfnamefont {S.}~\bibnamefont {Chattopadhyay}}, \bibinfo {author} {\bibfnamefont {G.~D.}\ \bibnamefont {Pace}}, \bibinfo {author} {\bibfnamefont {L.}~\bibnamefont {Skolc}}, \bibinfo {author} {\bibfnamefont {V.}~\bibnamefont {Helson}}, \bibinfo {author} {\bibfnamefont {S.}~\bibnamefont {Uchino}}, \bibinfo {author} {\bibfnamefont {E.}~\bibnamefont {Demler}}, \ and\ \bibinfo {author} {\bibfnamefont {J.-P.}\ \bibnamefont {Brantut}},\ }\href {https://arxiv.org/abs/2503.05420} {} (\bibinfo {year} {2025}),\ \Eprint {http://arxiv.org/abs/2503.05420} {arXiv:2503.05420} \BibitemShut {NoStop}%
\bibitem [{\citenamefont {Schlawin}\ and\ \citenamefont {Jaksch}(2019)}]{Schlawin2019atoms}%
  \BibitemOpen
  \bibfield  {author} {\bibinfo {author} {\bibfnamefont {F.}~\bibnamefont {Schlawin}}\ and\ \bibinfo {author} {\bibfnamefont {D.}~\bibnamefont {Jaksch}},\ }\bibfield  {title} {\bibinfo {title} {Cavity-mediated unconventional pairing in ultracold fermionic atoms},\ }\href {\doibase 10.1103/PhysRevLett.123.133601} {\bibfield  {journal} {\bibinfo  {journal} {Phys. Rev. Lett.}\ }\textbf {\bibinfo {volume} {123}},\ \bibinfo {pages} {133601} (\bibinfo {year} {2019})}\BibitemShut {NoStop}%
\bibitem [{\citenamefont {Ortu{\~n}o-Gonzalez}\ \emph {et~al.}(2025)\citenamefont {Ortu{\~n}o-Gonzalez}, \citenamefont {Lin}, \citenamefont {Stefaniak}, \citenamefont {Baumg{\"a}rtner}, \citenamefont {Natale}, \citenamefont {Donner},\ and\ \citenamefont {Chitra}}]{ortuno2025pauli}%
  \BibitemOpen
  \bibfield  {author} {\bibinfo {author} {\bibfnamefont {D.}~\bibnamefont {Ortu{\~n}o-Gonzalez}}, \bibinfo {author} {\bibfnamefont {R.}~\bibnamefont {Lin}}, \bibinfo {author} {\bibfnamefont {J.}~\bibnamefont {Stefaniak}}, \bibinfo {author} {\bibfnamefont {A.}~\bibnamefont {Baumg{\"a}rtner}}, \bibinfo {author} {\bibfnamefont {G.}~\bibnamefont {Natale}}, \bibinfo {author} {\bibfnamefont {T.}~\bibnamefont {Donner}}, \ and\ \bibinfo {author} {\bibfnamefont {R.}~\bibnamefont {Chitra}},\ }\bibfield  {title} {\bibinfo {title} {Pauli crystal superradiance},\ }\href {https://arxiv.org/abs/2505.02837} {\bibfield  {journal} {\bibinfo  {journal} {arXiv preprint arXiv:2505.02837}\ } (\bibinfo {year} {2025})}\BibitemShut {NoStop}%
\bibitem [{\citenamefont {Frank}\ \emph {et~al.}(2025)\citenamefont {Frank}, \citenamefont {Pini}, \citenamefont {Lang},\ and\ \citenamefont {Piazza}}]{frank2025fate}%
  \BibitemOpen
  \bibfield  {author} {\bibinfo {author} {\bibfnamefont {B.}~\bibnamefont {Frank}}, \bibinfo {author} {\bibfnamefont {M.}~\bibnamefont {Pini}}, \bibinfo {author} {\bibfnamefont {J.}~\bibnamefont {Lang}}, \ and\ \bibinfo {author} {\bibfnamefont {F.}~\bibnamefont {Piazza}},\ }\bibfield  {title} {\bibinfo {title} {The fate of the fermi surface coupled to a single-wave-vector cavity mode},\ }\href {https://arxiv.org/abs/2505.11452} {\bibfield  {journal} {\bibinfo  {journal} {arXiv preprint arXiv:2505.11452}\ } (\bibinfo {year} {2025})}\BibitemShut {NoStop}%
\bibitem [{\citenamefont {Veshchunov}\ \emph {et~al.}(2016)\citenamefont {Veshchunov}, \citenamefont {Magrini}, \citenamefont {Mironov}, \citenamefont {Godin}, \citenamefont {Trebbia}, \citenamefont {Buzdin}, \citenamefont {Tamarat},\ and\ \citenamefont {Lounis}}]{veshchunov2016optical}%
  \BibitemOpen
  \bibfield  {author} {\bibinfo {author} {\bibfnamefont {I.~S.}\ \bibnamefont {Veshchunov}}, \bibinfo {author} {\bibfnamefont {W.}~\bibnamefont {Magrini}}, \bibinfo {author} {\bibfnamefont {S.}~\bibnamefont {Mironov}}, \bibinfo {author} {\bibfnamefont {A.}~\bibnamefont {Godin}}, \bibinfo {author} {\bibfnamefont {J.-B.}\ \bibnamefont {Trebbia}}, \bibinfo {author} {\bibfnamefont {A.~I.}\ \bibnamefont {Buzdin}}, \bibinfo {author} {\bibfnamefont {P.}~\bibnamefont {Tamarat}}, \ and\ \bibinfo {author} {\bibfnamefont {B.}~\bibnamefont {Lounis}},\ }\bibfield  {title} {\bibinfo {title} {Optical manipulation of single flux quanta},\ }\href {https://www.nature.com/articles/ncomms12801} {\bibfield  {journal} {\bibinfo  {journal} {Nature communications}\ }\textbf {\bibinfo {volume} {7}},\ \bibinfo {pages} {12801} (\bibinfo {year} {2016})}\BibitemShut {NoStop}%
\bibitem [{\citenamefont {Sekiguchi}\ \emph {et~al.}(2024)\citenamefont {Sekiguchi}, \citenamefont {Narita}, \citenamefont {Hirori}, \citenamefont {Ono},\ and\ \citenamefont {Kanemitsu}}]{sekiguchi2024anomalous}%
  \BibitemOpen
  \bibfield  {author} {\bibinfo {author} {\bibfnamefont {F.}~\bibnamefont {Sekiguchi}}, \bibinfo {author} {\bibfnamefont {H.}~\bibnamefont {Narita}}, \bibinfo {author} {\bibfnamefont {H.}~\bibnamefont {Hirori}}, \bibinfo {author} {\bibfnamefont {T.}~\bibnamefont {Ono}}, \ and\ \bibinfo {author} {\bibfnamefont {Y.}~\bibnamefont {Kanemitsu}},\ }\bibfield  {title} {\bibinfo {title} {Anomalous behavior of critical current in a superconducting film triggered by dc plus terahertz current},\ }\href {https://www.nature.com/articles/s41467-024-48738-8} {\bibfield  {journal} {\bibinfo  {journal} {Nature Communications}\ }\textbf {\bibinfo {volume} {15}},\ \bibinfo {pages} {4435} (\bibinfo {year} {2024})}\BibitemShut {NoStop}%
\bibitem [{\citenamefont {De~Vecchi}\ \emph {et~al.}(2025)\citenamefont {De~Vecchi}, \citenamefont {Jotzu}, \citenamefont {Buzzi}, \citenamefont {Fava}, \citenamefont {Gebert}, \citenamefont {Fechner}, \citenamefont {Kimel},\ and\ \citenamefont {Cavalleri}}]{de2025generation}%
  \BibitemOpen
  \bibfield  {author} {\bibinfo {author} {\bibfnamefont {G.}~\bibnamefont {De~Vecchi}}, \bibinfo {author} {\bibfnamefont {G.}~\bibnamefont {Jotzu}}, \bibinfo {author} {\bibfnamefont {M.}~\bibnamefont {Buzzi}}, \bibinfo {author} {\bibfnamefont {S.}~\bibnamefont {Fava}}, \bibinfo {author} {\bibfnamefont {T.}~\bibnamefont {Gebert}}, \bibinfo {author} {\bibfnamefont {M.}~\bibnamefont {Fechner}}, \bibinfo {author} {\bibfnamefont {A.}~\bibnamefont {Kimel}}, \ and\ \bibinfo {author} {\bibfnamefont {A.}~\bibnamefont {Cavalleri}},\ }\bibfield  {title} {\bibinfo {title} {Generation of ultrafast magnetic steps for coherent control},\ }\href {https://www.nature.com/articles/s41566-025-01651-y} {\bibfield  {journal} {\bibinfo  {journal} {Nature Photonics}\ ,\ \bibinfo {pages} {1}} (\bibinfo {year} {2025})}\BibitemShut {NoStop}%
\bibitem [{\citenamefont {Van~Otterlo}\ \emph {et~al.}(1999)\citenamefont {Van~Otterlo}, \citenamefont {Golubev}, \citenamefont {Zaikin},\ and\ \citenamefont {Blatter}}]{VanOtterlo_1999}%
  \BibitemOpen
  \bibfield  {author} {\bibinfo {author} {\bibfnamefont {A.}~\bibnamefont {Van~Otterlo}}, \bibinfo {author} {\bibfnamefont {D.}~\bibnamefont {Golubev}}, \bibinfo {author} {\bibfnamefont {A.}~\bibnamefont {Zaikin}}, \ and\ \bibinfo {author} {\bibfnamefont {G.}~\bibnamefont {Blatter}},\ }\bibfield  {title} {\bibinfo {title} {Dynamics and effective actions of {{BCS}} superconductors},\ }\href {\doibase 10.1007/s100510050836} {\bibfield  {journal} {\bibinfo  {journal} {The European Physical Journal B}\ }\textbf {\bibinfo {volume} {10}},\ \bibinfo {pages} {131} (\bibinfo {year} {1999})}\BibitemShut {NoStop}%
\bibitem [{\citenamefont {Halperin}\ \emph {et~al.}(1974)\citenamefont {Halperin}, \citenamefont {Lubensky},\ and\ \citenamefont {Ma}}]{PhysRevLett.32.292}%
  \BibitemOpen
  \bibfield  {author} {\bibinfo {author} {\bibfnamefont {B.~I.}\ \bibnamefont {Halperin}}, \bibinfo {author} {\bibfnamefont {T.~C.}\ \bibnamefont {Lubensky}}, \ and\ \bibinfo {author} {\bibfnamefont {S.-k.}\ \bibnamefont {Ma}},\ }\bibfield  {title} {\bibinfo {title} {First-order phase transitions in superconductors and smectic-$a$ liquid crystals},\ }\href {\doibase 10.1103/PhysRevLett.32.292} {\bibfield  {journal} {\bibinfo  {journal} {Phys. Rev. Lett.}\ }\textbf {\bibinfo {volume} {32}},\ \bibinfo {pages} {292} (\bibinfo {year} {1974})}\BibitemShut {NoStop}%
\bibitem [{\citenamefont {Amelio}\ \emph {et~al.}(2021)\citenamefont {Amelio}, \citenamefont {Korosec}, \citenamefont {Carusotto},\ and\ \citenamefont {Mazza}}]{PhysRevB.104.235120}%
  \BibitemOpen
  \bibfield  {author} {\bibinfo {author} {\bibfnamefont {I.}~\bibnamefont {Amelio}}, \bibinfo {author} {\bibfnamefont {L.}~\bibnamefont {Korosec}}, \bibinfo {author} {\bibfnamefont {I.}~\bibnamefont {Carusotto}}, \ and\ \bibinfo {author} {\bibfnamefont {G.}~\bibnamefont {Mazza}},\ }\bibfield  {title} {\bibinfo {title} {Optical dressing of the electronic response of two-dimensional semiconductors in quantum and classical descriptions of cavity electrodynamics},\ }\href {\doibase 10.1103/PhysRevB.104.235120} {\bibfield  {journal} {\bibinfo  {journal} {Phys. Rev. B}\ }\textbf {\bibinfo {volume} {104}},\ \bibinfo {pages} {235120} (\bibinfo {year} {2021})}\BibitemShut {NoStop}%
\bibitem [{\citenamefont {Kakazu}\ and\ \citenamefont {Kim}(1994)}]{PhysRevA.50.1830}%
  \BibitemOpen
  \bibfield  {author} {\bibinfo {author} {\bibfnamefont {K.}~\bibnamefont {Kakazu}}\ and\ \bibinfo {author} {\bibfnamefont {Y.~S.}\ \bibnamefont {Kim}},\ }\bibfield  {title} {\bibinfo {title} {Quantization of electromagnetic fields in cavities and spontaneous emission},\ }\href {\doibase 10.1103/PhysRevA.50.1830} {\bibfield  {journal} {\bibinfo  {journal} {Phys. Rev. A}\ }\textbf {\bibinfo {volume} {50}},\ \bibinfo {pages} {1830} (\bibinfo {year} {1994})}\BibitemShut {NoStop}%
\bibitem [{\citenamefont {Pearl}(1964)}]{10.1063/1.1754056}%
  \BibitemOpen
  \bibfield  {author} {\bibinfo {author} {\bibfnamefont {J.}~\bibnamefont {Pearl}},\ }\bibfield  {title} {\bibinfo {title} {Current distribution in superconducting films carrying quantized fluxoids},\ }\href {\doibase 10.1063/1.1754056} {\bibfield  {journal} {\bibinfo  {journal} {Applied Physics Letters}\ }\textbf {\bibinfo {volume} {5}},\ \bibinfo {pages} {65} (\bibinfo {year} {1964})}\BibitemShut {NoStop}%
\bibitem [{\citenamefont {Altland}\ and\ \citenamefont {Simons}(2010)}]{altland2010condensed}%
  \BibitemOpen
  \bibfield  {author} {\bibinfo {author} {\bibfnamefont {A.}~\bibnamefont {Altland}}\ and\ \bibinfo {author} {\bibfnamefont {B.~D.}\ \bibnamefont {Simons}},\ }\href@noop {} {\emph {\bibinfo {title} {Condensed matter field theory}}}\ (\bibinfo  {publisher} {Cambridge university press},\ \bibinfo {year} {2010})\BibitemShut {NoStop}%
\bibitem [{\citenamefont {Li}\ \emph {et~al.}(2022)\citenamefont {Li}, \citenamefont {Schamri\ss{}},\ and\ \citenamefont {Eckstein}}]{PhysRevB.105.165121}%
  \BibitemOpen
  \bibfield  {author} {\bibinfo {author} {\bibfnamefont {J.}~\bibnamefont {Li}}, \bibinfo {author} {\bibfnamefont {L.}~\bibnamefont {Schamri\ss{}}}, \ and\ \bibinfo {author} {\bibfnamefont {M.}~\bibnamefont {Eckstein}},\ }\bibfield  {title} {\bibinfo {title} {Effective theory of lattice electrons strongly coupled to quantum electromagnetic fields},\ }\href {\doibase 10.1103/PhysRevB.105.165121} {\bibfield  {journal} {\bibinfo  {journal} {Phys. Rev. B}\ }\textbf {\bibinfo {volume} {105}},\ \bibinfo {pages} {165121} (\bibinfo {year} {2022})}\BibitemShut {NoStop}%
\bibitem [{Note1()}]{Note1}%
  \BibitemOpen
  \bibinfo {note} {This contrasts with the effect of purely thermal fluctuations studied in Ref. \cite {PhysRevLett.32.292}. There, the transverse EM field acquires a gap induced by the Meissner effect and couples directly to the mean-field $\Delta $ in the GL action. Since the Meissner gap vanishes near $T_c$, these fluctuations become very strong and, once integrated out, reshape the free energy profile and alter the nature of the phase transition. This effect is intrinsic and observable in type-I superconductors, but in our case it is rendered negligible by the cavity-induced photonic gap, that remains finite also at $T_c$}\BibitemShut {NoStop}%
\bibitem [{Note2()}]{Note2}%
  \BibitemOpen
  \bibinfo {note} {Strictly speaking, one must account for the more general and nonlocal Pippard regime $D^{-1}_\protect \text {cl}({\protect \bf q}) = Q({\protect \bf q})+{\protect \bf q}^2$, which exists for $\protect \tilde \kappa _\protect \text {GL}(\Sigma _0) \lesssim 1/\protect \sqrt {2}$. However, in the dirty case with small scattering length $\ell \ll \xi $ the Pippard kernel $Q({\protect \bf q}) \rightarrow \protect \tilde {\lambda }_L^{-2}$ and we restore the local electrodynamics.}\BibitemShut {Stop}%
\bibitem [{\citenamefont {Robbins}\ \emph {et~al.}(2025)\citenamefont {Robbins}, \citenamefont {Sedai}, \citenamefont {Howzen}, \citenamefont {Klaes}, \citenamefont {Loloee}, \citenamefont {Birge},\ and\ \citenamefont {Satchell}}]{robbins2025upper}%
  \BibitemOpen
  \bibfield  {author} {\bibinfo {author} {\bibfnamefont {K.~B.}\ \bibnamefont {Robbins}}, \bibinfo {author} {\bibfnamefont {P.}~\bibnamefont {Sedai}}, \bibinfo {author} {\bibfnamefont {A.~J.}\ \bibnamefont {Howzen}}, \bibinfo {author} {\bibfnamefont {R.~M.}\ \bibnamefont {Klaes}}, \bibinfo {author} {\bibfnamefont {R.}~\bibnamefont {Loloee}}, \bibinfo {author} {\bibfnamefont {N.~O.}\ \bibnamefont {Birge}}, \ and\ \bibinfo {author} {\bibfnamefont {N.}~\bibnamefont {Satchell}},\ }\bibfield  {title} {\bibinfo {title} {Upper critical fields in normal metal--superconductor--normal metal trilayers},\ }\href {https://www.nature.com/articles/s41598-025-98332-1} {\bibfield  {journal} {\bibinfo  {journal} {Scientific Reports}\ }\textbf {\bibinfo {volume} {15}},\ \bibinfo {pages} {13076} (\bibinfo {year} {2025})}\BibitemShut {NoStop}%
\bibitem [{\citenamefont {Jarc}\ \emph {et~al.}(2022)\citenamefont {Jarc}, \citenamefont {Mathengattil}, \citenamefont {Giusti}, \citenamefont {Barnaba}, \citenamefont {Singh}, \citenamefont {Montanaro}, \citenamefont {Glerean}, \citenamefont {Rigoni}, \citenamefont {Zilio}, \citenamefont {Winnerl},\ and\ \citenamefont {Fausti}}]{10.1063/5.0080045}%
  \BibitemOpen
  \bibfield  {author} {\bibinfo {author} {\bibfnamefont {G.}~\bibnamefont {Jarc}}, \bibinfo {author} {\bibfnamefont {S.~Y.}\ \bibnamefont {Mathengattil}}, \bibinfo {author} {\bibfnamefont {F.}~\bibnamefont {Giusti}}, \bibinfo {author} {\bibfnamefont {M.}~\bibnamefont {Barnaba}}, \bibinfo {author} {\bibfnamefont {A.}~\bibnamefont {Singh}}, \bibinfo {author} {\bibfnamefont {A.}~\bibnamefont {Montanaro}}, \bibinfo {author} {\bibfnamefont {F.}~\bibnamefont {Glerean}}, \bibinfo {author} {\bibfnamefont {E.~M.}\ \bibnamefont {Rigoni}}, \bibinfo {author} {\bibfnamefont {S.~D.}\ \bibnamefont {Zilio}}, \bibinfo {author} {\bibfnamefont {S.}~\bibnamefont {Winnerl}}, \ and\ \bibinfo {author} {\bibfnamefont {D.}~\bibnamefont {Fausti}},\ }\bibfield  {title} {\bibinfo {title} {Tunable cryogenic terahertz cavity for strong light–matter coupling in complex materials},\ }\href {\doibase 10.1063/5.0080045} {\bibfield  {journal} {\bibinfo  {journal} {Review of Scientific Instruments}\ }\textbf {\bibinfo {volume} {93}},\ \bibinfo {pages} {033102} (\bibinfo {year} {2022})}\BibitemShut {NoStop}%
\bibitem [{\citenamefont {Tolpygo}(2016)}]{10.1063/1.4948618}%
  \BibitemOpen
  \bibfield  {author} {\bibinfo {author} {\bibfnamefont {S.~K.}\ \bibnamefont {Tolpygo}},\ }\bibfield  {title} {\bibinfo {title} {Superconductor digital electronics: Scalability and energy efficiency issues (review article)},\ }\href {\doibase 10.1063/1.4948618} {\bibfield  {journal} {\bibinfo  {journal} {Low Temperature Physics}\ }\textbf {\bibinfo {volume} {42}},\ \bibinfo {pages} {361} (\bibinfo {year} {2016})}\BibitemShut {NoStop}%
\bibitem [{\citenamefont {Gao}\ \emph {et~al.}(2020)\citenamefont {Gao}, \citenamefont {Schlawin}, \citenamefont {Buzzi}, \citenamefont {Cavalleri},\ and\ \citenamefont {Jaksch}}]{gao2020photoinduced}%
  \BibitemOpen
  \bibfield  {author} {\bibinfo {author} {\bibfnamefont {H.}~\bibnamefont {Gao}}, \bibinfo {author} {\bibfnamefont {F.}~\bibnamefont {Schlawin}}, \bibinfo {author} {\bibfnamefont {M.}~\bibnamefont {Buzzi}}, \bibinfo {author} {\bibfnamefont {A.}~\bibnamefont {Cavalleri}}, \ and\ \bibinfo {author} {\bibfnamefont {D.}~\bibnamefont {Jaksch}},\ }\bibfield  {title} {\bibinfo {title} {Photoinduced electron pairing in a driven cavity},\ }\href {\doibase https://doi.org/10.1103/PhysRevLett.125.053602} {\bibfield  {journal} {\bibinfo  {journal} {Phys. Rev. Lett.}\ }\textbf {\bibinfo {volume} {125}},\ \bibinfo {pages} {053602} (\bibinfo {year} {2020})}\BibitemShut {NoStop}%
\bibitem [{\citenamefont {Chiocchetta}\ \emph {et~al.}(2021)\citenamefont {Chiocchetta}, \citenamefont {Kiese}, \citenamefont {Zelle}, \citenamefont {Piazza},\ and\ \citenamefont {Diehl}}]{chiocchetta2021cavity}%
  \BibitemOpen
  \bibfield  {author} {\bibinfo {author} {\bibfnamefont {A.}~\bibnamefont {Chiocchetta}}, \bibinfo {author} {\bibfnamefont {D.}~\bibnamefont {Kiese}}, \bibinfo {author} {\bibfnamefont {C.~P.}\ \bibnamefont {Zelle}}, \bibinfo {author} {\bibfnamefont {F.}~\bibnamefont {Piazza}}, \ and\ \bibinfo {author} {\bibfnamefont {S.}~\bibnamefont {Diehl}},\ }\bibfield  {title} {\bibinfo {title} {Cavity-induced quantum spin liquids},\ }\href {\doibase https://doi.org/10.1038/s41467-021-26076-3} {\bibfield  {journal} {\bibinfo  {journal} {Nature Communications}\ }\textbf {\bibinfo {volume} {12}},\ \bibinfo {pages} {5901} (\bibinfo {year} {2021})}\BibitemShut {NoStop}%
\end{thebibliography}
\end{document}